\documentclass[aps,prx,onecolumn,preprint,amsfonts,nobibnotes,superscriptaddress,showpacs, longbibliography]{revtex4-1}

\usepackage{dcolumn}
\usepackage{amsmath}
\usepackage{amssymb}
\usepackage{graphicx}
\usepackage{bm}
\usepackage{color}
\usepackage{titlesec}

\begin{document}

\title{Clocking the Onset of Bilayer Coherence in a High-$\bm{\mathrm{T_C}}$ Cuprate}

\vspace{2cm}

\author{Edoardo Baldini}
	\affiliation{Laboratory for Ultrafast Microscopy and Electron Scattering, IPHYS, EPFL, CH-1015 Lausanne, Switzerland}
	\affiliation{Laboratory of Ultrafast Spectroscopy, ISIC, EPFL, CH-1015 Lausanne, Switzerland}

\author{Andreas Mann}
	\affiliation{Laboratory for Ultrafast Microscopy and Electron Scattering, IPHYS, EPFL, CH-1015 Lausanne, Switzerland}

\author{Benjamin P. P. Mallett}
	\affiliation{Department of Physics, University of Fribourg, Chemin du Mus\'ee 3, CH-1700 Fribourg, Switzerland}

\author{Christopher Arrell}
	\affiliation{Laboratory of Ultrafast Spectroscopy, ISIC, EPFL, CH-1015 Lausanne, Switzerland}
	
\author{Frank van Mourik}
	\affiliation{Laboratory of Ultrafast Spectroscopy, ISIC, EPFL, CH-1015 Lausanne, Switzerland}

\author{Thomas Wolf}
	\affiliation{Institute of Solid State Physics, Karlsruhe Institute of Technology, Postfach 3640, Karlsruhe 76021, Germany}
	
\author{Dragan Mihailovic}
	\affiliation{Jozef Stefan Institute and International Postgraduate School, Jamova 39, SI-1000 Ljubljana, Slovenia}
	
\author{Jeffrey L. Tallon}
	\affiliation{Robinson Research Institute, Victoria University of Wellington, P.O. Box 33436 Lower Hutt, New Zealand}
	
\author{Christian Bernhard}
	\affiliation{Department of Physics, University of Fribourg, Chemin du Mus\'ee 3, CH-1700 Fribourg, Switzerland}

\author{Jos\'e Lorenzana}
	\affiliation{Institute for Complex Systems - CNR, and Physics Department, University of Rome "La Sapienza", I-00185 Rome, Italy}

\author{Fabrizio Carbone}
	\affiliation{Laboratory for Ultrafast Microscopy and Electron Scattering, IPHYS, EPFL, CH-1015 Lausanne, Switzerland}

\date{\today}

\begin{abstract}
In cuprates, a precursor state of superconductivity is speculated to exist above the critical temperature $\mathrm{T_C}$. Here we show via a combination of far-infrared ellipsometry and ultrafast broadband optical spectroscopy that signatures of such a state can be obtained via three independent observables in an underdoped sample of NdBa$_2$Cu$_3$O$_{6+\delta}$. The pseudogap correlations were disentangled from the response of laser-broken pairs by clocking their characteristic time-scales. The onset of a superconducting precursor state was found at a temperature $\mathrm{T_{ONS}}$ $>$ $\mathrm{T_C}$, consistent with the temperature scale identified via static optical spectroscopy. Furthermore, the temperature evolution of the coherent vibration of the Ba ion, strongly renormalized by the onset of superconductivity, revealed a pronounced anomaly at the same temperature $\mathrm{T_{ONS}}$. The microscopic nature of such a precursor state is discussed in terms of pre-formed pairs and enhanced bilayer coherence.
\end{abstract}

\pacs{}

\maketitle

\section{Introduction}

The short coherence length ($\xi_0\sim 1$ nm) of Cooper pairs in cuprate high-$\mathrm{T_C}$ superconductors allows for a variety of fascinating phenomena in contrast to low-$\mathrm{T_C}$ materials, which have homogeneous superconducting (SC) properties on length scales $\xi_0$ of the order of several hundreds or thousands of nanometers~\cite{ref:Tinkham}. On one hand, superconductivity and other electronic states can coexist in cuprates, with disorder tipping the balance on a local scale. On the other hand, thermal and quantum fluctuations of the SC order parameter can play an important role because of the reduced dimensionality (layered structure) of the material. Thus, understanding the interplay between SC fluctuations, inhomogeneities, competing orders and reduced dimensionality remains a major challenge in cuprate physics.

\noindent Several temperature scales have been identified in the cuprates phase diagram (Fig. 1(a)), which presumably result from the above interplay, but whose precise meaning is far from being understood. Below $\mathrm{T^*}$, the pseudogap (PG) state appears (Fig. 1(a), grey circles)~\cite{Warren1989,Alloul1989,Takigawa1991,Loram1993}. 
Early ideas~\cite{ref:emery, ref:kosztin} suggested that the PG reflected the presence of pairing correlations without long-range phase coherence. Another line of thought postulates the existence of a different kind of incipient electronic order~\cite{Zaanen1989,Emery1993,Castellani1995,Benfatto2000,Nayak2000,ref:chubukov,Varma1999,Capati2015}
competing with superconductivity. Traces of such orders have been seen in different regions of the phase diagram, as stripes~\cite{Tranquada1995,Bianconi1996}, nematic~\cite{Hinkov2008}, time-reversal symmetry breaking~\cite{Fauque2006} and incommensurate charge-density-waves (CDW)~\cite{ref:vershinin,Ghiringhelli2012,ref:hashimoto}.

\noindent More recent experiments give support to the competing scenario, by showing a temperature scale for precursor effects with a doping dependence quite different from $\mathrm{T^*}$. For example, scanning tunneling microscopy (STM) reveals that local pairing correlations can be detected up to several tenths of kelvin above the $\mathrm{T_C}$ at optimal doping (Fig. 1(a), brown squares)~\cite{ref:gomes,Yazdani2009}. Angle-resolved photoemission spectroscopy provides a similar temperature scale~\cite{ref:kondo1,ref:kondo2}. In addition, local probes show the inhomogeneous nature of the phenomena~\cite{Iguchi2001,ref:gomes,Yazdani2009}. Nernst effect~\cite{ref:wang} (Fig. 1(a), violet triangles) and magnetization~\cite{Li2010b} (Fig. 1(a), violet diamonds) measurements show another crossover line where precursor diamagnetic effects appear, requiring some degree of intralayer coherence.

\noindent In the case of bilayer materials, the far-infrared (FIR) \textit{c}-axis conductivity provides additional information. The response can be well described by a multilayer model of coupled bilayers separated by poorly conducting regions~\cite{VanderMarel1996,ref:munzar,ref:dubroka}. In this case, precursor effects appear as an increase in the Drude spectral weight (SW) due to the coherent transport between neighboring layers. Such a bilayer coherence requires substantial intralayer coherence to set in first, and indeed it appears closer to $\mathrm{T_C}$ (Fig. 1(a), red circles). For example, in optimally doped (OP) materials, the onset of the bilayer coherence coincides with $\mathrm{T_C}$ and only in the underdoped (UD) samples the bilayer onset temperature ($\mathrm{T_{ONS}}$) separates from $\mathrm{T_C}$, remaining always clearly below the PG temperature $\mathrm{T^*}$~\cite{ref:dubroka, ref:uykur}.
  
A powerful strategy for disentangling the above interplay is to separate the different contributions directly in real time via pump-probe spectroscopy~\cite{ref:kaindl, ref:smallwood, ref:hu, ref:mankowsky, ref:giannetti_review}. This technique allows to perturb the equilibrium between different states with a pump pulse and to subsequently study their incoherent recovery time or the dynamics of coherent modes linked to the perturbed states. In this regard, special attention has been reserved for the transient optical response of cuprate superconductors in the near-infrared/visible spectral range~\cite{ref:han,ref:albrecht,ref:thomas,ref:cjstevens,ref:mihailovic,ref:smith,ref:demsar, ref:mihailovic_Physica,ref:gedik_reflectivity,ref:giannetti,ref:torchinsky,ref:mansart,ref:Hinton2013,ref:madan, ref:fausti}. In OP cuprates, pump-probe spectroscopy provides a simple picture, as shown in Fig. 1(b), where we collect a number of single-wavelength transient reflectivity ($\Delta$R/R) data below $\mathrm{T_C}$ (solid lines) and above $\mathrm{T_C}$ (dashed lines) on a variety of OP materials under comparable experimental conditions~\cite{ref:torchinsky, ref:giannetti, ref:smith, ref:mihailovic_Physica}. In the normal-state above $\mathrm{T_C}$, a fast relaxation of several hundreds of fs appears. This is typically attributed to the cooling of a hot quasi-equilibrium electron gas, giving rise to a response similar to that observed in metals (the PG signal does not manifest in OP cuprates, see Fig. 1(a)). Below $\mathrm{T_C}$, the dynamics is instead dominated by a slow relaxation component $\tau_{QP}$ of several picoseconds, attributed to the recombination of quasiparticles (QPs) into pairs. Indeed, the relaxation dynamics of this QP response is very well described by the Rothwarf-Taylor model \cite{ref:gedik_reflectivity, ref:kabanov} and are directly related to the recovery of the SC gap detected by nonequilibrium low-energy probes~\cite{ref:kaindl, ref:saichu, ref:pashkin, ref:cortes, ref:smallwood, ref:graf, ref:wzhang, ref:wzhang2}. That said, one should be aware that the Rothwarf-Taylor model was derived for long coherence length superconductors and the possibility to disorder the phase of preformed pairs in cuprates may bring new physics.

\begin{figure}[t]
\centering
\includegraphics[width=0.6\columnwidth]{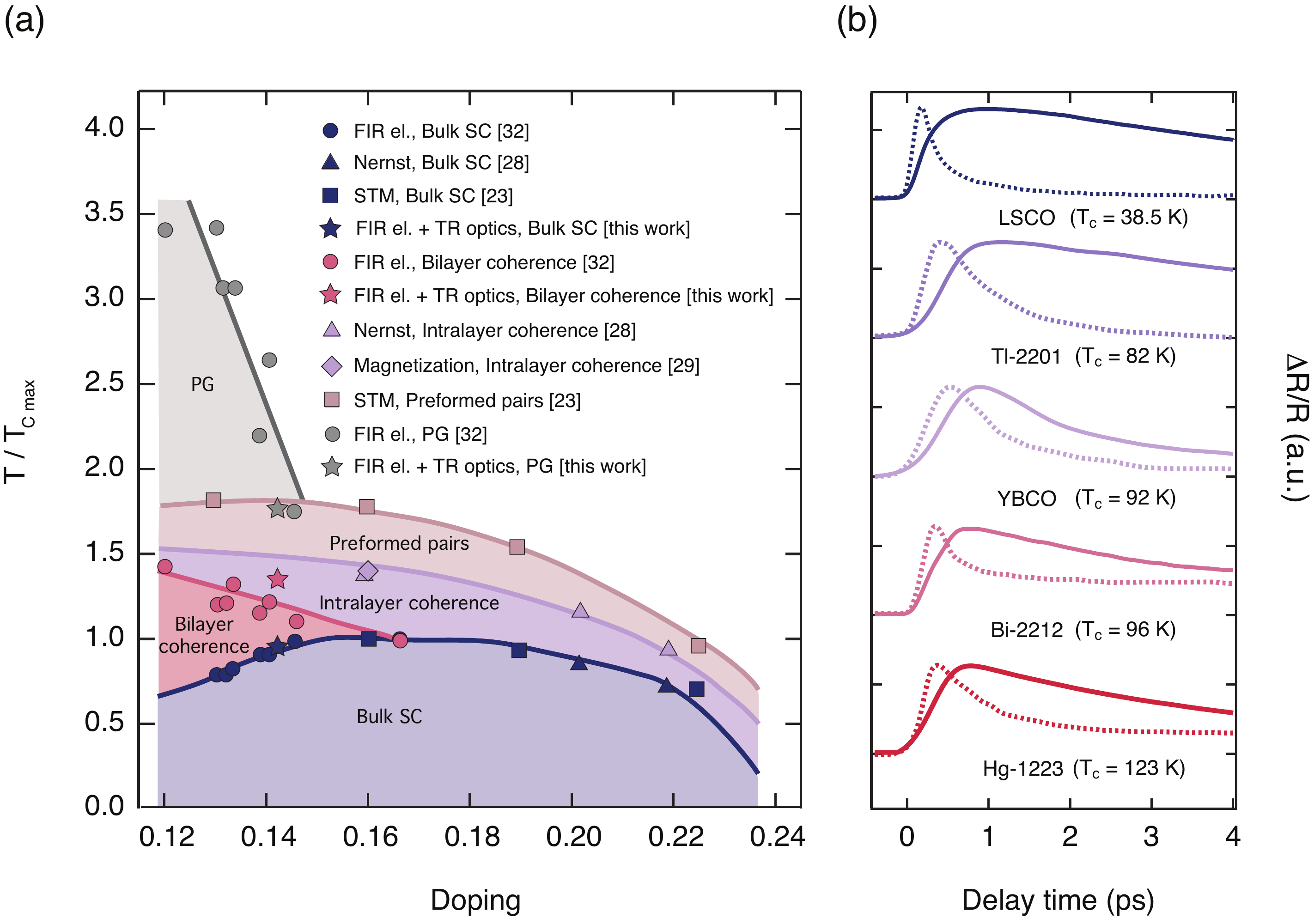}
\caption{(a)
Phase diagram for bilayer cuprates displaying the temperature scale for SC fluctuations. The data points have been obtained by a number of experimental studies: STM~\cite{ref:gomes}, Nernst effect~\cite{ref:wang}, magnetization~\cite{Li2010b}, FIR ellipsometry~\cite{ref:dubroka} and this work. Symbols and colors are highlighted in the label. (b) Normalized single-wavelength transient reflectivity traces collected in the SC (solid lines) and normal (dashed lines) states on different OP cuprates under comparable experimental conditions. Data have been adapted from Refs. ~\cite{ref:torchinsky, ref:giannetti, ref:smith, ref:mihailovic_Physica}. The critical temperature for each material is indicated in the figure.}
\end{figure}
 
In UD samples a more complex behaviour occurs, due to the emergence of a multi-component response. Indeed, the QP recombination dynamics across the SC gap is accompanied by an additional fast decay time $\tau_{PG}$ lasting several hundreds of femtoseconds. Since this component vanishes at $\mathrm{T^*}$, it has been ascribed to the recombination of the carriers subjected to PG correlations. Finally, superimposed to these relaxations, a very long decay time $\mathrm{\tau_H}$ of several ns appears, usually interpreted either as a pump-induced heating effect~\cite{ref:han,ref:mihailovic,ref:smith,ref:gedik_reflectivity,ref:giannetti, ref:madan} or as the signature of a photoinduced absorption from localized carriers~\cite{ref:thomas, ref:cjstevens, ref:mihailovic_Physica}. Pioneering experiments also suggested the possibility to detect precursor effects within the PG phase of UD cuprates, by monitoring the evolution of the QP signal above $\mathrm{T_C}$~\cite{ref:mihailovic}. In contrast to OP cuprates, in UD samples this component persists well above $\mathrm{T_C}$.

\noindent Another strength of pump-probe optical spectroscopy consists in the possibility to reveal the ultrafast dynamics and the intrinsic properties of specific Raman-active bosonic collective modes~\cite{ref:albrecht, ref:mansart, ref:torchinsky, ref:Hinton2013, ref:fausti, ref:mann}, which are coherently excited via the Impulsive Stimulated Raman Scattering process or by a long-lived perturbation of the electronic ground state~\cite{ref:testevens}. Interestingly, several of these modes have strong intensity and energy anomalies at $\mathrm{T_C}$, which suggest that they can be used as probes of pairing correlations~\cite{ref:albrecht, ref:fausti, ref:Hinton2013}. Similar anomalies are seen in spontaneous Raman scattering~\cite{ref:friedl}.

\begin{figure*}[t]
\centering
\includegraphics[height=5.3cm]{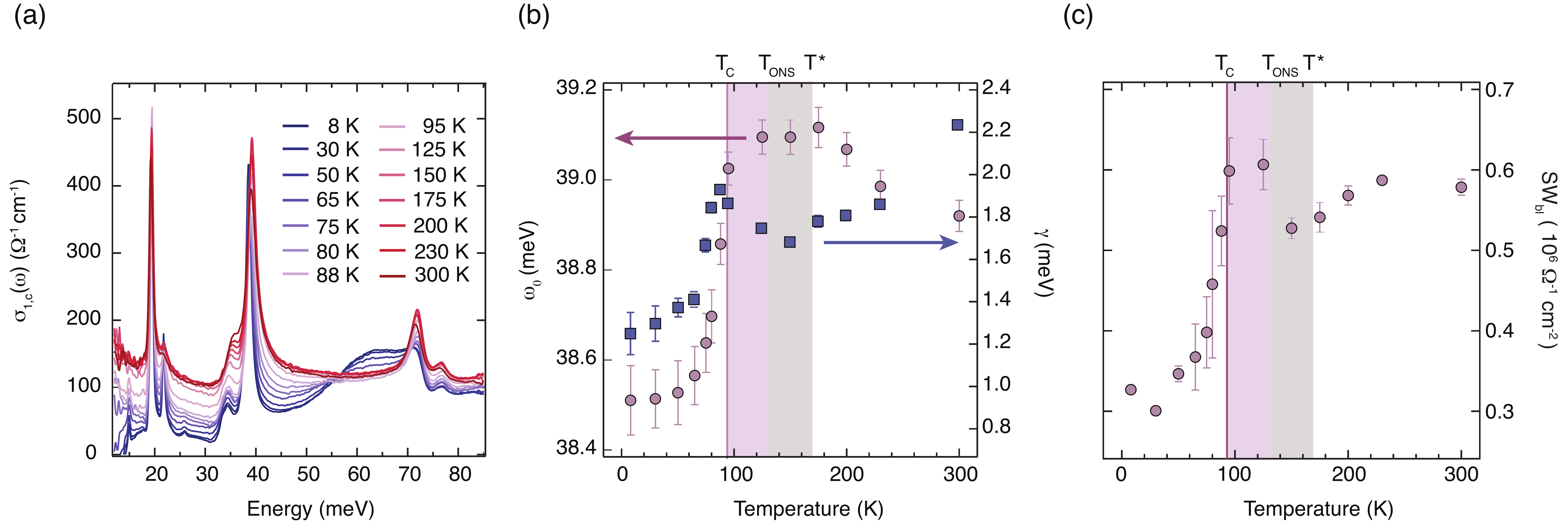}
\caption{(a) Real part of the \textit{c}-axis optical conductivity in the FIR measured by ellipsometry as a function of temperature. (b) Temperature evolution of the $\mathrm{B_{1u}}$ O bending mode frequency (violet circles) and linewidth (blue squares), showing an anomalous behavior between $\mathrm{T_C}$ and $\mathrm{T_{ONS}}$. The violet vertical line marks the value of $\mathrm{T_C}$ and the violet (grey) shaded area highlights the temperature region characterized by incoherent pairing (PG) correlations. (c) Temperature dependence of the low-energy SW of the bilayer conductivity.}
\end{figure*}

All these results make ultrafast spectroscopy a suitable candidate to study the dynamics of precursor SC effects. Here, we perform a combination of far-infrared (FIR) ellipsometry and femtosecond broadband optical spectroscopy as a function of temperature to identify different spectroscopic features associated with the precursor SC state in an UD cuprate. As a model system, we select a high-quality single-crystal of slightly UD NdBa$_2$Cu$_3$O$_{6+\delta}$ (NBCO), which is isostructural to YBCO and has a sufficiently large difference between $\mathrm{T_C}$ (93.5 K) and $\mathrm{T^*}$ (170 K) while being close to optimal doping~\cite{ref:yu}. Therefore, it is the ideal playground for identifying the spectral fingerprint of the QP signal above $\mathrm{T_C}$ and for testing how its coherent phonon modes react to the possible existence of precursor phenomena.

In the nonequilibrium experiment, we reveal the QP response over a large spectral range, showing that this signature persists well above $\mathrm{T_C}$ and vanishes only above a characteristic temperature $\mathrm{T_{ONS}}$ $\sim$ 130 K. Simultaneously, the anomaly of the coherent Ba mode is observed at $\mathrm{T_{ONS}}$ and not close to $\mathrm{T_C}$ as in OP compounds. These findings are supported by the steady-state optical data in the FIR, which provide an indipendent estimate of $\mathrm{T_{ONS}}$ by means of a local electric field-analysis and via the temperature dependence of the $\mathrm{B_{1u}}$ O bending mode~\cite{ref:munzar, ref:dubroka}. Our data suggest that a precursor SC state is present at these temperature, where bilayer coherence is established among planes containing pre-formed pairs.

\section{Results}

\subsection{Far-infrared ellipsometry} 
In order to locate our sample in the phase diagram of Fig. 1(a), we measure the \textit{c}-axis optical conductivity by means of FIR spectroscopic ellipsometry. This quantity is very sensitive to the opening of a gap in the density of states, especially near the antinodal region of the Fermi-surface where a PG develops below $\mathrm{T^*}$ in UD samples~\cite{ref:yu, ref:andersen}. This is due to the strong k-dependence of the perpendicular hopping matrix element for the transfer of charge carriers between the CuO$_2$ bilayer units (across the BaO and CuO chain layers). The formation of the SC gap below $\mathrm{T_C}$ gives rise to an additional weaker suppression of the conductivity and to a pronounced mode at finite frequency (transverse Josephon plasma mode)~\cite{ref:yu}. The latter arises from the layered structure which is composed of two closely spaced CuO$_2$ layer (bilayer unit) and a spacer layer that consists of two BaO layers and the CuO chain layer. This transverse plasma mode is very sensitive to the coherence of the electron transport between the CuO$_2$ layers of the individual bilayer units and thus to the onset of SC correlations, even if they are short-ranged and strongly fluctuating.

\noindent The data obtained on our NBCO single-crystal are reported in Fig. 2(a) as a function of temperature. As a first step, we identify the PG and SC gap temperature scales of our sample via the analysis of the low-energy SW. The details are reported in the Supplemental Material (SM)~\cite{ref:yu}. This analysis yields $\mathrm{T_C}$ = 94 K and $\mathrm{T^*}$ = 170 K. 

\noindent Subsequently, we estimate the $\mathrm{T_{ONS}}$ temperature scale by applying the same analysis performed in Ref.~\cite{ref:dubroka}. In particular, we focus on the central frequency and linewidth in the optical response of the $\mathrm{B_{1u}}$ phonon at $\sim$ 39 meV. This corresponds to the bending mode of the O ions in the CuO$_2$ plane and it is also affected by the onset of the bilayer coherence which modifies the local fields on the CuO$_2$ layers~\cite{ref:munzar, ref:dubroka, ref:uykur}. Fig. 2(b) shows the temperature evolution of the phonon peak position and linewidth revealing a renormalization in the temperature range between 110 K and 140 K that signifies the onset of the bilayer coherence between these temperatures.

\noindent The $\mathrm{T_{ONS}}$ temperature scale is also estimated by a local electric field analysis of the data using the so-called multilayer model~\cite{VanderMarel1996, ref:munzar, ref:dubroka}. Within this model, $\mathrm{T_{ONS}}$ is associated with the temperature below which the bilayer conductivity starts to exhibit an increase~\cite{ref:dubroka, ref:corson}. For $\mathrm{T_{C}}$$ < $T$ < $$\mathrm{T_{ONS}}$, the condensation of pairs with a finite correlation lifetime enhances the coherence among the neighboring planes of a bilayer and thus gives rise to an increase in the low-energy SW. In the model, the real and imaginary parts of the conductivity are simultaneously fitted using the multi-layer model in the energy range from 16 meV to 370 meV. The parameters describing the phonons are fitted at 300 K and essentially kept fixed at all lower temperatures, allowing only for a small reduction in the width and a blueshift of the peak energy due to thermal effects. Figure 2(c) shows the SW of the low-energy component of the bilayer conductivity. There is a significant increase in the low-energy SW below 150 K, before it decreases again below 95 K due to the opening of a full SC gap (which is accompanied by a transfer of SW into a $\delta$ function at zero energy). The resulting estimate of $\mathrm{T_{ONS}}$, between 110 K and 140 K, is therefore consistent with the value extracted from the B$_{1u}$ phonon analysis.

\begin{figure}[t]
\centering
\includegraphics[width=0.4\columnwidth]{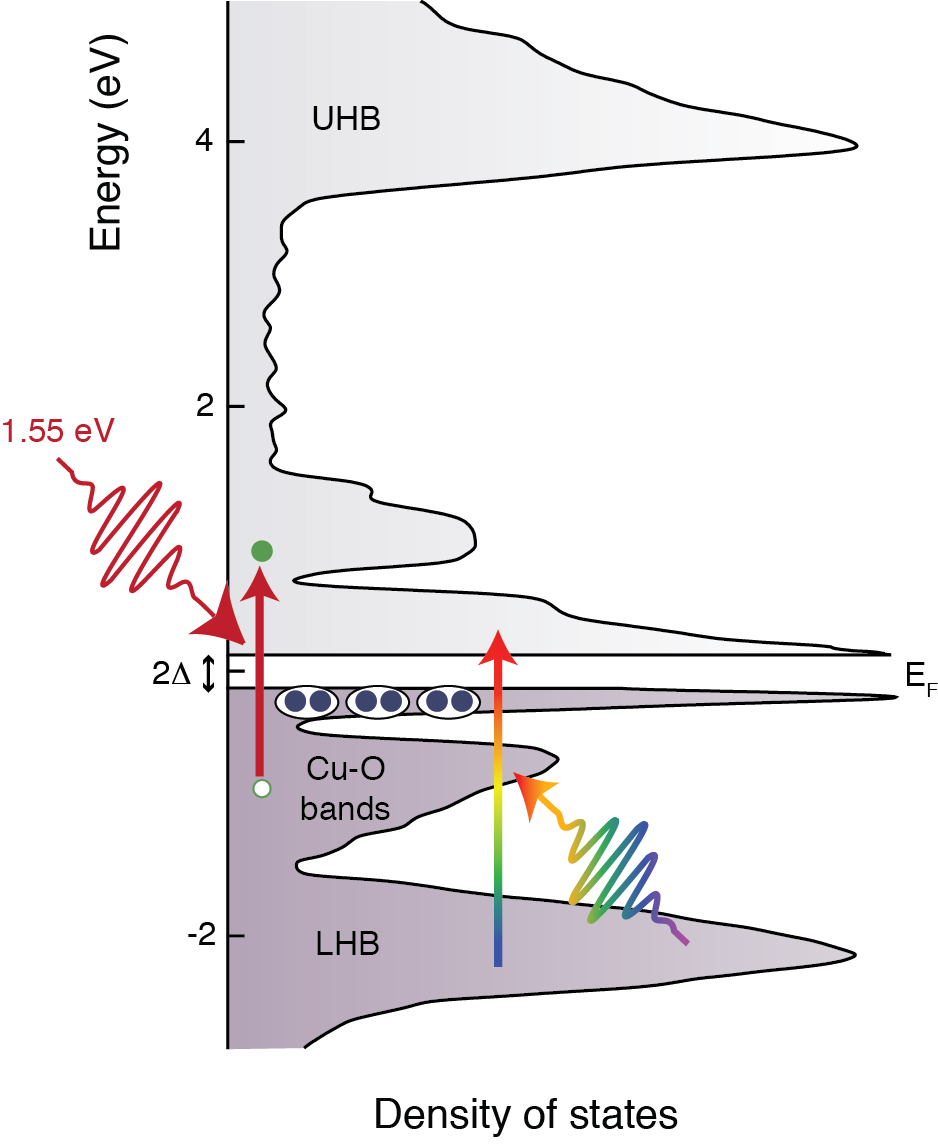}
\caption{Schematic illustration of the density of states for the 123-family of cuprates. The density of states is adapted from Dynamical Mean Field Theory calculations~\cite{ref:fausti}. The pump photon energy at 1.55 eV is depicted in dark red and the white light continuum probe is represented with a rainbow arrow. The photogenerated particle-hole pairs across $\mathrm{E_F}$ are displayed in green, the pairs close to $\mathrm{E_F}$ in blue. UHB = Upper Hubbard Band, LHB = Lower Hubbard Band.}
\end{figure}

\noindent The three temperature scales obtained for our NBCO single-crystal are displayed by star symbols in the phase diagram of Fig. 1(a). The slight shifts of our values with respect to those reported in the phase diagram can be associated with a different scaling shown by NBCO single-crystals compared to YBCO.

\subsection{Time-resolved broadband reflectivity}
The ability of femtosecond light excitation to set superconductivity out of equilibrium offers another route for disentangling the different temperature scales. In our time-resolved optical study, we tune the pump photon energy to be resonant with an interband charge excitation that promotes particle-hole pairs across $\mathrm{E_F}$ towards high-energy states~\cite{ref:mihailovic, ref:fausti}. Afterwards, we monitor the optical reflectivity change $\Delta$R/R in a broad spectral region between 1.70 and 2.80 eV. This spectral range includes the absorption feature at 2.60 eV, involving occupied states in the Lower Hubbard Band and unoccupied states close to $\mathrm{E_F}$~\cite{ref:fausti}. A pictorial illustration is offered in Fig. 3, where the one-particle density of states computed by Dynamical Mean Field Theory (adapted from ref. \cite{ref:fausti}) is shown together with the optical transitions promoted by our pump and probe pulses. Details about the steady-state optical conductivity are reported in SM, while the description of the pump-probe experiment is given in the Methods section.

\begin{figure*}[t]
\centering
\includegraphics[height=9cm]{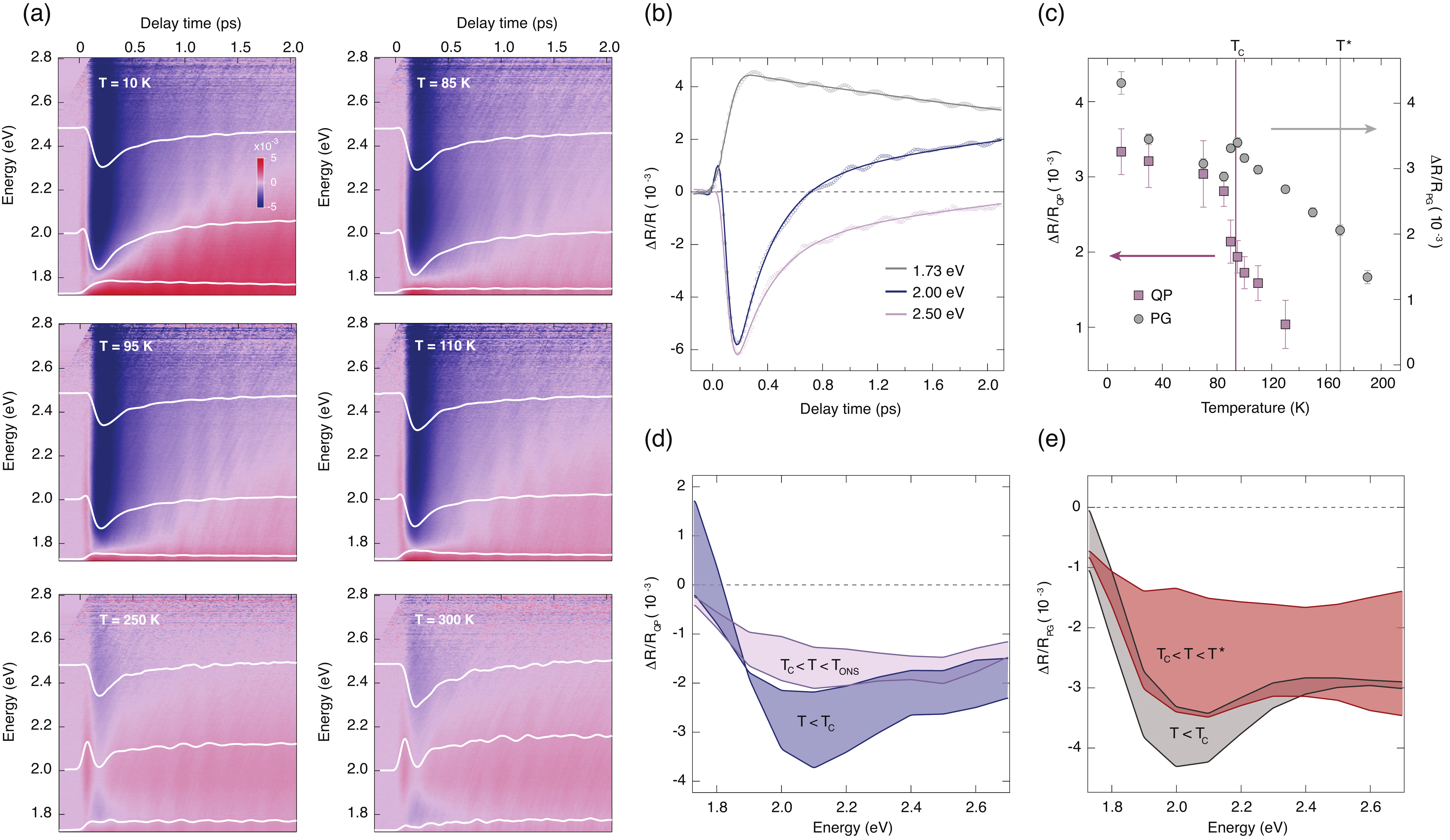}
\caption{(a) Transient reflectivity $\Delta$R/R($\tau$, E) as a function of the probe photon energy and time delay between pump and probe. The pump photon energy is set at 1.55 eV and the absorbed fluence corresponds to $\sim$ 300 $\mu$J/cm$^2$. The temperatures are indicated in the labels. (b) Temporal traces at \mbox{10 K} for fixed probe photon energies of 1.73 eV, 2.00 eV and 2.50 eV are shown as dotted lines. The results of the global fit analysis are reported on top as solid lines. (c) Temperature evolution of the QP (violet squares) and PG (grey circles) contributions to $\Delta$R/R at 2.00 eV probe photon energy, as extracted from the global fit analysis. The vertical violet and grey lines identify the temperature scales $\mathrm{T_C}$ and $\mathrm{T^*}$, respectively. (d, e) Transient reflectivity spectra of the QP and PG response obtained from the global fit analysis. Different colour shadings are used to identify the temperature regions below $\mathrm{T_C}$ (blue), and between $\mathrm{T_C}$ and $\mathrm{T_{ONS}}$ (violet) for the QP response or below $\mathrm{T_C}$ (grey), and between $\mathrm{T_{C}}$ and $\mathrm{T^*}$ (red) for the PG response.}
\end{figure*}

\noindent Figure 4(a) displays $\Delta$R/R as a function of the probe photon energy and time delay between pump and probe for some representative temperatures (10, 85, 95, 110, 250 and 300 K). The temporal dynamics of $\Delta$R/R at 10 K are displayed in Fig. 4(b) for selected probe photon energies (1.73 eV, 2.00 eV and 2.50 eV). The signal around 2.00 eV comprises a first, resolution-limited positive rise, followed by a second, delayed negative contribution that shows the maximum change of the response around 200 fs. Subsequently, a fast decay of the negative contribution takes place and the response becomes positive at larger time delays. This pattern characterizes the signal across the whole spectrum, but the relative weights of the positive and negative components are strongly dependent on the probe photon energy. On top of this incoherent response an oscillation can be seen in the whole spectrum and can be ascribed to the coherent excitation of collective bosonic modes. We measure $\Delta$R/R of our sample at 16 temperatures. The full set of data is included in the SM. 

\subsection{Global fit analysis}
As a first step in our analysis, we perform a global fit of $\Delta$R/R as a function of time in order to disentangle the contributions of the different processes occurring on our ultrafast timescale. Eleven temporal traces are selected from each map of the temperature dependence and fitted simultaneously by imposing the same time constants. At low-temperatures, a satisfactory fit of the incoherent signal can only be obtained by using three exponential functions convolved with a Gaussian response accounting for the temporal shape of the pump pulse. Whereas a resolution-limited rise characterizes the first exponential term, a delayed rise has to be used for the other two components to correctly reproduce the early dynamics. More details of the fitting procedure are provided in the SM. The time constants governing the three components match the ones reported in literature \cite{ref:han,ref:mihailovic,ref:smith, ref:mihailovic_Physica,ref:giannetti,ref:kaindl, ref:pashkin, ref:madan}, corresponding to the PG response $\tau_{PG}$ (hundreds of fs), the QP response $\tau_{QP}$ (several ps) and the long-lived component $\tau_{H}$ (hundreds of ps). The results of the fit at 10 K are reported in Fig. 4(b) as solid lines superimposed on the experimental traces. 
When iterated for all the measured $\Delta$R/R maps, the global fit analysis allows to track the separate temperature evolution of the QP and PG responses. In Fig. 4(c) we report their temperature dependences for a selected probe photon energy of 2.00 eV, at which detailed single-wavelength pump-probe studies on OP YBCO have been reported~\cite{ref:han, ref:albrecht}. Interestingly, in our UD sample, the presence of the QP response is detected even above $\mathrm{T_C}$, up to a temperature $\mathrm{T_{ONS}}$ $\sim$ 130 K, which we associate with the onset of a precursor SC state and which is consistent with the previous equilibrium analysis. The faster PG contribution instead sets in near $\mathrm{T^*}$ and increases its weight down to $\mathrm{T_C}$. 

\noindent In contrast to single-wavelength pump-probe studies, our broadband probe pulse retrieves the whole $\Delta$R/R spectrum of each contribution. The $\Delta$R/R spectrum of the QP response obtained from the global fit is shown in Fig. 4(d), where, for clarity, we group the spectra below $\mathrm{T_C}$ and between $\mathrm{T_C}$ and $\mathrm{T_{ONS}}$ in different colour shadings. Figure 4(e) reports the spectrum of the PG component in the temperature ranges below $\mathrm{T_{C}}$ and between $\mathrm{T_{C}}$ and $\mathrm{T^*}$. By resolving the whole optical spectrum, we can observe that the QP response exhibits a sign reversal at 1.80 eV. It undergoes a continuous decrease in its amplitude with increasing temperature only in two spectral ranges, below 1.75 eV and around 2.00 - 2.10 eV. These results are fully consistent with previous experiments performed around 1.55 eV and 2.00 eV~\cite{ref:mihailovic}, and demonstrate that the temperature dependence of low-energy phenomena in cuprates can be easily tracked by these specific high-energy photons in a manner similar to low-energy probes~\cite{ref:pashkin, ref:kaindl}. Although the measured $\Delta$R/R spectra represent a mixture of the dispersive and absorptive parts of the system's dielectric function, its identification allows one to determine the transient optical conductivity response of the system, as shown below. This unique capability of broadband ultrafast optical spectroscopy enables us to overcome the limitations of single-wavelength studies and provides a deeper connection with the material's electronic structure.

\subsection{Transient optical conductivity}
As anticipated above, the use of a continuum probe can provide quantitative information on the microscopic processes affecting the visible spectral range of our cuprate. A useful quantity that can be extracted from the nonequilibrium experiment is the transient complex optical conductivity $\Delta\sigma$/$\sigma$ = $\Delta\sigma_1$/$\sigma_1$ + i $\Delta\sigma_2$/$\sigma_2$. This can be calculated without the need of a Kramers-Kronig transform by relying on steady-state spectroscopic ellipsometry data we measured in the visible range (reported in the SM) as a starting point and performing a Drude-Lorentz analysis of the $\Delta$R/R maps at the different temperatures. In particular, the determination of the real part $\Delta\sigma_1$/$\sigma_1$ gives access to the temporal evolution of the SW in the visible range.

\noindent In Fig. 5(a) we report $\Delta\sigma_1$/$\sigma_1$ at 10 K as a function of probe photon energy and time delay. A prominent drop is found in a wide energy range between 1.90 and 2.80 eV at early time delays. In contrast, for energies below 1.80 eV, a positive contribution emerges and progressively dominates the higher energy range. The response at early time delays strongly differs from the one expected from a simple transient heating of the crystal, which would display a positive sign across the whole measured spectrum (see the temperature dependence of $\sigma_1$ in the SM). Hence, we can already assume that the pump pulse predominantly acts on the crystal as a non-thermal perturbation, creating a nonequilibrium distribution of hot carriers across $\mathrm{E_F}$.

\begin{figure}[t]
\centering
\includegraphics[width=0.6\columnwidth]{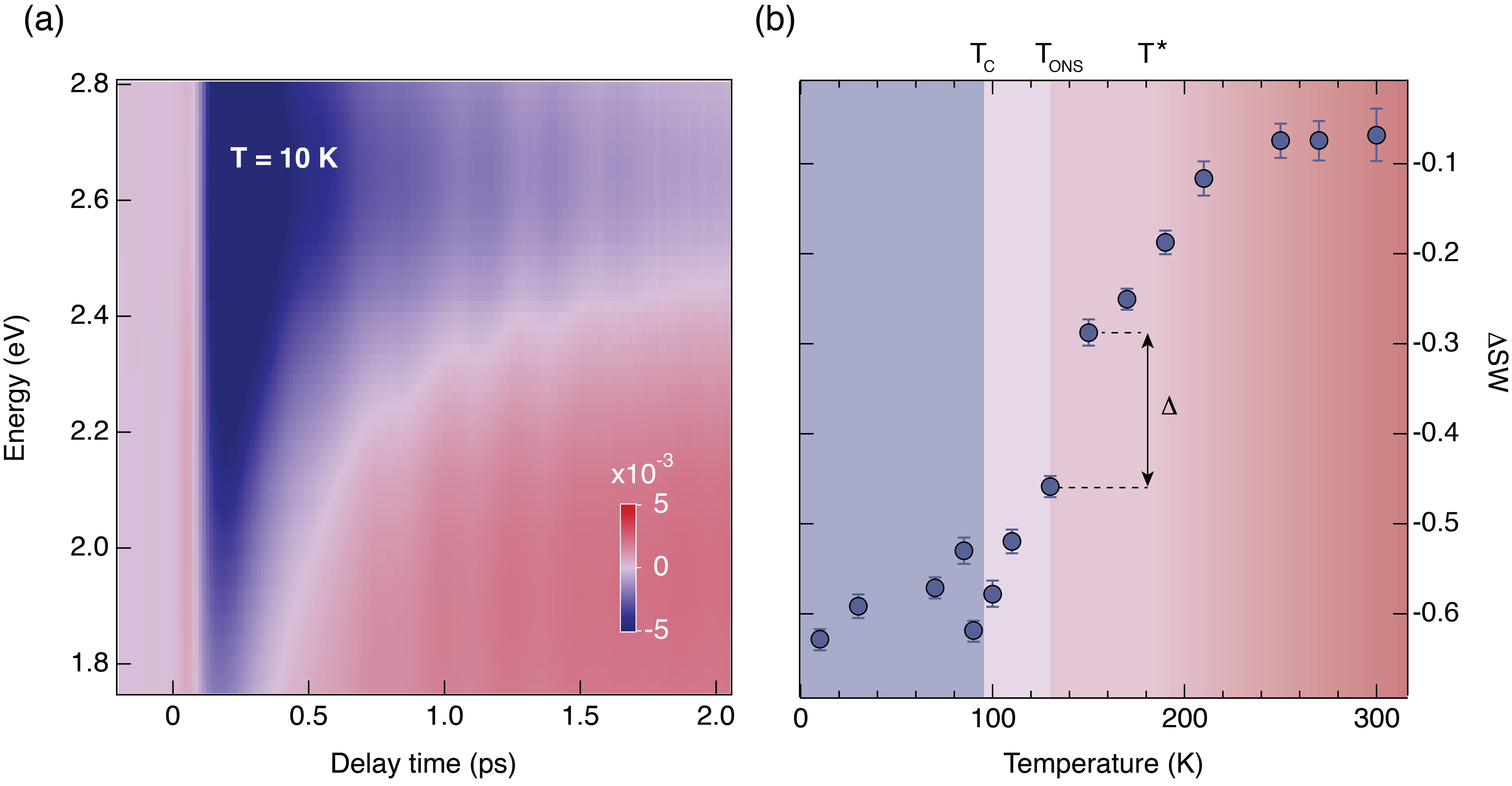}
\caption{(a) Calculated transient optical conductivity $\Delta\sigma_1/\sigma_1$ at 10 K as a function of probe photon energy and of time delay between pump and probe. (b) Temperature evolution of the nonequilibrium SW ($\Delta$SW) integrated over the whole probe spectrum at 200 fs delay time. The blue shaded region highlights the temperature range where the material is superconducting, the violet region depicts the preformed state and the gradient red region represents the crossover from the PG phase to the normal state of the material. The respective temperature scales $\mathrm{T_C}$, $\mathrm{T_{ONS}}$ and $\mathrm{T^*}$ are indicated on top. The $\Delta$ parameter is indicated on the graph and identifies the difference existing between the $\Delta$SW at 150 K and 130 K for a time delay of 200 fs.}
\end{figure}

\noindent To unravel the effects produced by each separate ultrafast process on $\Delta\sigma_1$/$\sigma_1$, we fit ten temporal traces with the same model function used for $\Delta$R/R. In this way the spectral dependences of the different components can be traced, as reported in the SM. The long-lived component of the response provides an increase of the SW that is mostly flat and can be reproduced from the steady-state $\sigma_1$ by assuming a simple heating of the crystal. In contrast, the QP and PG contributions are responsible for the pronounced decrease of the SW in the visible range (peaked around the interband transition at 2.70 eV) and manifest their maximum amplitude around 200 fs, suggesting the occurrence of a delayed response. This indicates that the photoinduced hot carriers decay within 200 fs to the proximity of $\mathrm{E_F}$. A fraction of the energy released in this fast relaxation is conveyed to the particles subjected to SC and PG correlations, providing a channel for their excitation. Since these excited particles accumulate close to $\mathrm{E_F}$, the final states associated with the high-energy interband transitions from the Lower Hubbard Band to $\mathrm{E_F}$ result occupied. This explains why the broadband probe experiences a reduced absorption in the visible range due to a Pauli blocking mechanism. The SW removed in the visible range is typically transferred to low energy, contributing significantly to the onset of a transient Drude response~\cite{ref:pashkin} and possibly opening new channels that were inactive under equilibrium conditions.

\noindent Finally, the determination of $\Delta\sigma_1$/$\sigma_1$ at all temperatures allows to follow the temperature evolution of the change in SW ($\Delta$SW) in the probed range, as displayed in Fig. 5(b) for a time delay of 200 fs. Remarkably, $\Delta$SW displays its maximum absolute intensity in the SC state and shows a first kink in the proximity $\mathrm{T_C}$, which can be related to the opening of the SC gap. As the temperature is increased above $\mathrm{T_C}$, $\Delta$SW reduces its value, which is indicative for the occurrence of a rapid crossover. A pronounced recovery of $\Delta$SW is found around $\mathrm{T_{ONS}}$, as the QP response from the precursor state ceases to contribute to the loss of SW in the probed spectral range. Finally, as the temperature approaches $\mathrm{T^*}$, a second crossover takes place and the response stabilizes around a vanishingly small constant value. Thus, by selecting the difference between $\Delta$SW($\tau$ = 200 fs) at 150 K and 130 K, our experiment measures the amount of SW ``lost'' in the visible range due to the existence of QPs in the precursor state. This quantity, which we name $\Delta$, also contains a minor contribution from the particles subjected to PG correlations, since the amplitude of their nonequilibrium response varies (although slightly) with temperature. Recently, it has been shown that three-pulse experiments can directly disentangle these two contributions in time without the need of a subtraction procedure~\cite{ref:madan}. In this scheme, a first pulse is used to selectively melt the SC condensate, and the recovery dynamics is measured via a conventional two-pulse pump-probe. This technique has been demonstrated using a single colour for the three pulses, but we expect that its future extension to a broadband detection can contribute to refine the current value of $\Delta$. However, already from our analysis of the incoherent response, we can conclude that signatures of the precursor state above $\mathrm{T_C}$ emerge in the nonequilibrium experiment.

\subsection{Coherent phonon analysis}
The above analysis is complemented by the temperature dependence of the Raman-active modes which are coherently excited by the pump pulse and give rise to an oscillatory modulation across the probed spectral region. An important advantage of our technique over single-wavelength studies is that the broadband probe gives access to the energy dependent Raman matrix elements of all collective modes affecting the visible range. When corroborated by theory, this methodology provides a very selective and quantitative estimate of the electron-phonon coupling~\cite{ref:mann, ref:fausti}. Moreover, the shape of the Raman matrix elements enables us to track the temperature dependence of the coherent modes in a spectral region where they resonate with specific electronic excitations and can be clearly distinguished. To assign the collective modes present in our spectra, we perform a Fourier transform (FT) analysis of the residuals from the global fit. With this approach, we identify the presence of two separate modes influencing the high-energy electrodynamics of our UD cuprate. The Raman matrix elements of the two separate coherent modes across the probed range are shown and discussed in Fig. S13. Here, in Fig. 6(a), we report only the FT spectra around the probe photon energy of 2.10 eV at the selected temperatures of 10 K, 110 K, 130 K and 300 K, which clearly reveal the presence of the two separate peaks. The choice of this probe photon energy lies in the high visibility of both modes around this spectral region, as suggested by the Raman matrix elements shape.

\begin{figure*}[t]
\centering
\includegraphics[height=6cm]{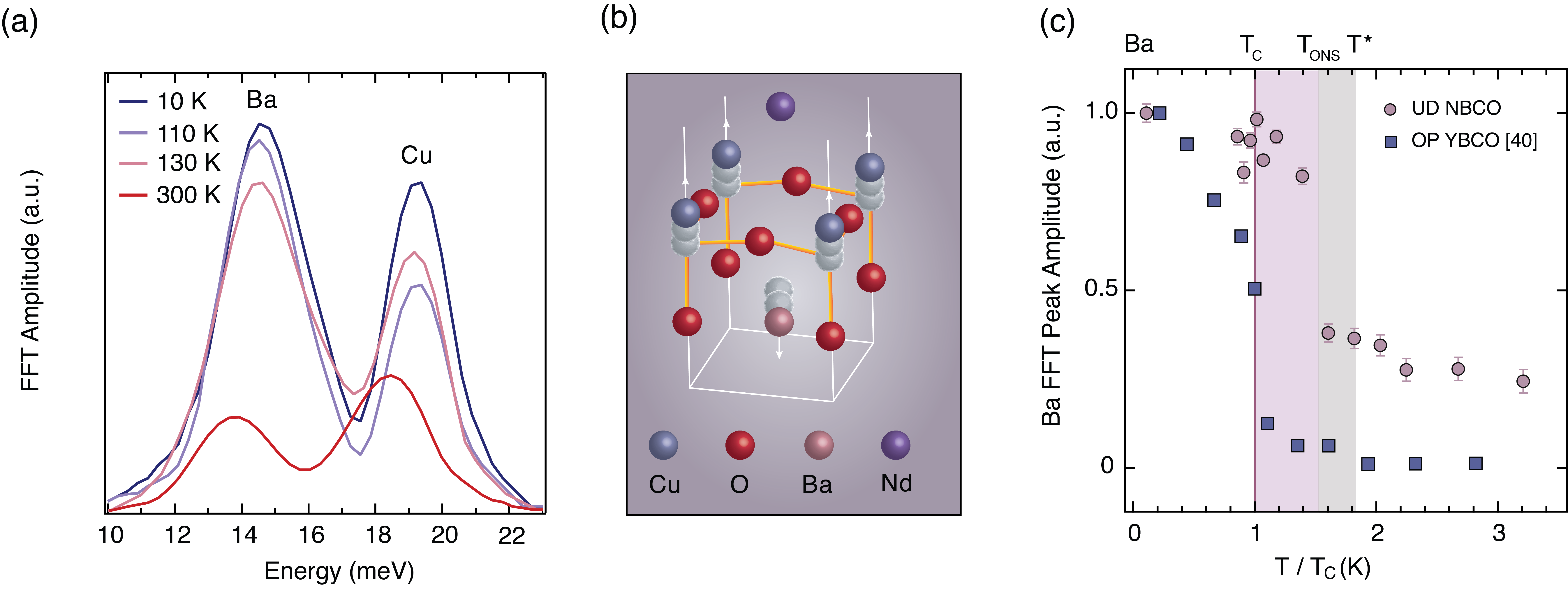}
\caption{(a) Fourier transform analysis of the residuals obtained from the global fit at selected temperatures. The two modes at $\sim$ 15 meV and $\sim$ 19 meV correspond to the coherent $\mathrm{A_{1g}}$ Ba and Cu phonons, respectively. (b) Atomic displacements of the Raman-active $\mathrm{A_{1g}}$ phonons involving c-axis vibrations of the Ba and Cu ions. The down-pointing arrow corresponds to the Ba mode, while the up-pointing arrows correspond to the Cu mode. (c) Temperature dependence of the Ba mode amplitude at 2.10 eV of probe photon energy in OP YBCO (blue squares) and UD NBCO (violet circles). The data for OP YBCO have been extracted from Ref. ~\cite{ref:albrecht}. The violet vertical line identifies the $\mathrm{T_C}$ temperature scale, while the violet (grey) shaded area highlights the temperature region of NBCO which is subjected to incoherent pairing (PG) correlations.}
\end{figure*}

\noindent The energies of the two peaks are $\sim$14.5 meV and $\sim$19.5 meV at 10 K and undergo a gradual softening as the temperature is increased. The peaks can be assigned to the totally-symmetric ($\mathrm{A_{1g}}$) Raman-active phonons involving the c-axis vibrations of the Ba and Cu ions, respectively. The atomic displacements are shown in Fig. 6(b). These modes have been extensively discussed in spontaneous Raman scattering and time-resolved pump-probe spectroscopy of OP YBCO~\cite{ref:fausti, ref:albrecht, ref:friedl, ref:misochko_timeresolved_YBCO_2000}. These two experimental techniques provide the same information concerning the frequency of the phonon modes, but they measure distinct states of the phonon system. Spontaneous Raman scattering measures an equilibrium response of the system. In pump-probe spectroscopy, the collective modes are instead brought into a coherent state by the pump pulse via an impulsive or displacive mechanism~\cite{ref:testevens} and therefore they may be influenced by the nonequilibirum distribution of carriers and by nonlinear effects.

\noindent In OP YBCO both experimental methods consistently observed: i) the softening of the Cu and Ba phonon frequencies for increasing temperatures and ii) an anomalous intensity increase of the Ba mode when the crystal entered the SC phase at $\mathrm{T_C}$. While in spontaneous Raman scattering the latter effect has been explained as a superconductivity-induced resonant process (in which the Raman cross-section is renormalized by self-energy effects), a different explanation has been provided by pump-probe spectroscopy: The anomalous temperature dependence of the Ba mode was demonstrated to follow the same behaviour of the QP response, suggesting that the driving force behind the coherent excitation is represented by the change in the density of broken pairs~\cite{ref:albrecht}. This displacive mechanism at the origin of the coherent Ba mode has also been invoked by theoretical calculations~\cite{ref:mazin}: In this scenario, the pair-breaking process instantaneously removes pairing energy from the energy balance determining the equilibrium ionic conditions, eventually triggering the coherent lattice motion. This explains why the coherent oscillation of the Ba mode in OP YBCO vanishes together with the QP signal as soon as $\mathrm{T_C}$ is crossed. Beyond such phenomenological models, recent \textit{ab initio} calculations found a high sensitivity of the density of states at $\mathrm{E_F}$ to the displacement of the Ba ion~\cite{ref:fausti}. Such strong electron-phonon coupling suggests that a perturbation of the low-energy electronic states can trigger the coherent motion of the Ba mode. Irrespective of the details of the excitation process, there is a firm experimental evidence that the Ba mode is a sensitive probe of the pairing correlations.


\noindent All the above observations were reported in OP YBCO, in which there is little or no coherence among the bilayers above $\mathrm{T_C}$ (Fig. 1(a)). In UD samples of the 123 family, the presence of bilayer coherence above $\mathrm{T_C}$ is therefore expected to play a major role in modifying the intensity anomaly of the Ba mode. Taking advantage of the determination of the energy dependent Raman matrix element, in our experiment we tracked the intensity of the Ba peak in the FT as a function of temperature. In Fig. 6(c) we show the results of this analysis on our slightly UD NBCO and we compare the temperature dependence of the normalized Ba mode intensity with the data reported in Ref.\cite{ref:albrecht} on OP YBCO. The data are displayed as a function of the normalized temperature $\mathrm{T}/\mathrm{T_C}$ and the temperature scales $\mathrm{T_{ONS}}$ and $\mathrm{T^*}$ for our slightly UD NBCO sample are highlighted.

\noindent Interestingly, in the UD crystal, we find that the Ba mode persists well above $\mathrm{T_C}$ and loses intensity only in the proximity of $\mathrm{T_{ONS}}$ (Fig. 6(c)). This anomaly indicates that the Ba mode intensity is sensitive to the stabilization of the bilayer coherence, as expected from the above discussion. Hence, we attribute the large intensity change displayed by the Ba mode to the establishment of the precursor SC state, consistent with the emergence of the QP component in the incoherent pump-probe response. An alternative possibility will also be proposed in the discussion.


\section{Discussion}
In our broadband ultrafast optical study we found evidence within three independent observables that suggest the emergence of a precursor SC state in the UD regime at a temperature scale $\mathrm{T_{ONS}}$, such that $\mathrm{T_C}$ $<$ $\mathrm{T_{ONS}}$ $<$ $\mathrm{T^*}$. In particular, these observables include the persistence of the QP response up to $\mathrm{T_{ONS}}$, a prominent jump in the QP SW in the vicinity of this temperature scale and the anomaly of the coherent Ba mode intensity at $\mathrm{T_{ONS}}$ and not at $\mathrm{T_C}$ as in OP cuprates. We remark that all these effects are simultaneously detected in the same transient spectra, providing a clear evidence for the development of an electronic state that affects the optical response of the system and that can be unambigously disentangled from the PG correlations on the basis of their distinct relaxation timescales. The interpretation of such an electronic state in terms of precursor SC correlations is based on the matching between the $\mathrm{T_{ONS}}$ scale under nonequilibrium conditions and the temperature extracted from FIR ellipsometry of the same UD single-crystal. Indeed, FIR ellipsometry demonstrates that the intra-bilayer response becomes more coherent below $\mathrm{T_{ONS}}$ and that this trend is just enhanced below $\mathrm{T_C}$. In the past, the application of a magnetic field has also been shown to counteract this trend, both below $\mathrm{T_C}$ and $\mathrm{T_{ONS}}$~\cite{ref:dubroka}. Although this effect is a clear evidence for enhanced SC correlations below $\mathrm{T_{ONS}}$, it does not exclude the possibility that the precursor SC state is coupled to another kind of fluctuating order, \textit{e.g.} CDW or spin-density wave~\cite{ref:dubroka}. This idea is reinforced when the portions of the YBCO phase diagram characterized by SC- and CDW-correlations are compared, revealing a remarkable correspondence between the $\mathrm{T_{ONS}}$ temperature scales associated with the two phenomena~\cite{ref:dubroka, ref:chang}. As an example, in YBCO Ortho-VIII ($\mathrm{T_C}$ = 67 K) the precursor SC state temperature $\mathrm{T_{ONS}}$ = 130 K measured by FIR ellipsometry~\cite{ref:dubroka} matches the temperature scale at which CDW has been detected by resonant elastic x-ray scattering~\cite{ref:chang}. Consistent with this scenario, recent single-wavelength pump-probe experiments proposed that the superfluid in UD cuprates can be considered as a condensate of coherently-mixed particle-particle and particle-hole pairs (particle-hole quadruplets), whose origin lies in a coupled SC-CDW order parameter~\cite{ref:hinton}. In the specific case of the 123 family of cuprates, one can also speculate that the coupling between the coherent Ba mode and the SC state is mediated by the competition among the latter and the CDW. As shown in Ref. ~\cite{ref:Hinton2013}, the repulsive interaction between the CDW and SC order parameters works as an efficient mechanism to establish the CDW in the SC state. Since the CDW is naturally a charged mode, it will couple strongly to all lattice phonons with the appropriate symmetry. Hence, once the CDW is established, coherent phonons will be also triggered due to their mutual coupling.

\noindent A complete understanding of this possibly intertwined precursor state setting at $\mathrm{T_{ONS}}$ and of its dynamical fluctuations will become possible only upon its direct imaging with sub-ps time resolution. In this perspective, the development of techniques as time-resolved STM and cryo-Lorentz microscopy is expected to play a crucial role~\cite{ref:terada, ref:cocker, ref:rajeswari}.

\begin{acknowledgments}
Work at LUMES was supported by NCCR MUST and by the ERC starting grant USED258697. B.P.P.M. and C.B. acknowledge the financial support of the Swiss National Science Foundation (SNSF) through Grant No. 200020-153660 and technical support during the ellipsometry experiments at the IR beam line of the ANKA synchrotron at FZ Karlsruhe, Germany.
\end{acknowledgments}
\newpage

\setcounter{section}{0}
\setcounter{figure}{0}
\renewcommand{\thesection}{S\arabic{section}}  
\renewcommand{\thetable}{S\arabic{table}}  
\renewcommand{\thefigure}{S\arabic{figure}} 
\renewcommand\Im{\operatorname{\mathfrak{Im}}}
\titleformat{\section}[block]{\bfseries}{\thesection.}{1em}{} 

\section{Methods}

\subsection{Sample growth and preparation.}
A high quality NdBa$_2$Cu$_3$O$_{6+x}$ (NBCO) single-crystal was flux-grown in an Y-stabilized zirconia crucible and under low oxygen partial pressure to avoid spurious substitution of the Nd ion onto the Ba site~\cite{ref:schlachter}. The resulting crystal is a parallelepiped with dimensions 5.5 mm $\times$ 4 mm $\times$ 1 mm along the \textit{a}, \textit{b} and \textit{c} axes respectively. The sample was annealed in oxygen for ten days at 370$^{\circ}$C resulting in a superconducting transition temperature of $\mathrm{T_{C}}$ = 93.5 K, as measured by dc magnetisation, and a sharp transition width of 1.5 K. The crystal surfaces were mechanically polished to optical grade using diamond powder paste.

\subsection{Broadband spectroscopic ellipsometry.}

Using broadband ellipsometry, we measured the $c$-axis and in-plane complex optical conductivity of the sample, covering the spectral range from 12.5 meV to 6.0 eV. Anisotropy corrections were performed using standard numerical procedures~\cite{ref:bernhard, ref:azzam} and diffraction effects at low frequency were accounted for using the procedure developed by Huml{\i}́{\v{c}}ek \textit{et al.}~\cite{ref:humlicek}. We used a home-built ellipsometer attached to a Bruker fast-Fourier spectrometer at the infrared beam line of the ANKA synchrotron at the Karlsruhe Institute for Technology to measure from 12.5 meV to 85.0 meV \cite{ref:bernhard} and a Woollam VASE ellipsometer at the University of Fribourg for 0.5 eV to 6.0 eV. When at cryogenic temperatures, the latter measurements were performed in a vacuum better than 10$^{-8}$ mbar to prevent measurable ice-condensation onto the sample.

\subsection{Ultrafast broadband optical spectroscopy.}

Femtosecond broadband transient reflectivity experiments have been performed using a set-up described in detail in Ref.~\cite{ref:mann}. A cryogenically-cooled amplified laser system provided sub-50 fs pulses centred around 1.55 eV at a repetition rate of 6 kHz. One third of the output, representing the probe beam, was focused on a CaF$_2$ cell to generate broadband visible pulses with a bandwidth covering the spectral range between 1.70 eV and 2.80 eV. The probe beam was then collimated and focused onto the sample through a pair of parabolic mirrors. The remaining two thirds of the output of the amplifier, representing the pump beam, were directed towards the sample under normal incidence. Along the pump path a chopper with a 60 slot plate was inserted, operating at 1.5 kHz and phase-locked to the laser-system. The polarizations for both the pump and the probe beam were set along the [110] crystallographic direction, giving access to the in-plane optical response of the sample and selecting the $\mathrm{A_{1g} + B_{2g}}$ Raman symmetry configuration. Pump and probe were focused onto the sample with spatial dimensions of \mbox{120 $\mathrm{\mu m}$ $ \times$ 87 $\mathrm{\mu m}$} and 23 $\mathrm{\mu m}$ $\times$ 23 $\mathrm{\mu m}$, respectively. The sample was mounted inside the chamber of a closed cycle cryostat, which provides a temperature-controlled environment in the range 10 - 340 K. The measurements were performed in a vacuum better than 10$^{-7}$ mbar and several temperature cycles have been applied to prevent measurable ice-condensation onto the sample. The reflected probe beam was dispersed by a fiber-coupled 0.3 m spectrograph and detected on a shot-to-shot basis with a CMOS linear array.

\section{Far-infrared ellipsometry}

We estimate the $\mathrm{T_C}$ and $\mathrm{T^*}$ temperature scales from the analysis of the real ($\sigma_1$) part of the $c$-axis conductivity, as measured by steady-state ellipsometry in the far-infrared spectral region. The data are reported in Fig. 2(a) of the main text. For our purposes, in the present analysis, we only focus on the evaluation of the spectral weight (SW) in the 12.5 - 85.0 meV spectral region as a function of temperature. We calculate the SW as:

\begin{equation}
\mathrm{SW} = \underset{12.5 \textnormal{meV}}{\overset{85 \textnormal{meV}}{\mathop \int }}\,{\sigma_{1,c}(\omega)}d\omega
\end{equation}

\begin{figure}[h!]
\includegraphics[width=0.4\columnwidth]{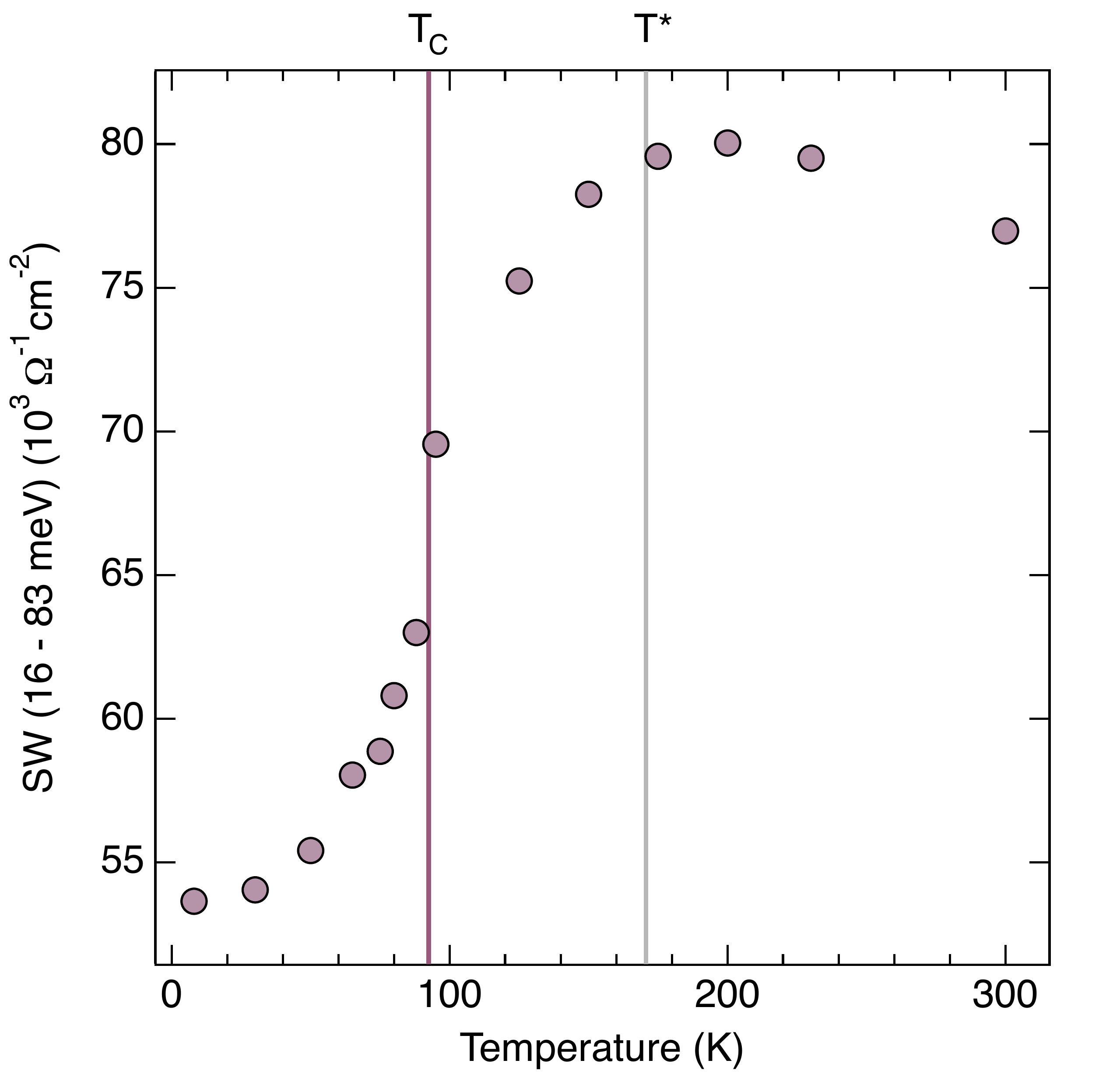}
\caption{Temperature dependence of the c-axis spectral weight, calculated over the far-infrared spectral range covered by spectroscopic ellipsometry (12.5 - 85.0 meV). The violet and grey vertical lines mark the estimated $\mathrm{T_C}$ and $\mathrm{T^*}$ respectively.}
\label{fig:FigureS1}
\end{figure}

In Fig. \ref{fig:FigureS1} the temperature dependence of the SW is shown. As expected, we observe that the SW decreases well above $\mathrm{T_C}$. This `missing' SW is a characteristic feature of the pseudogap \cite{ref:yu} and provides an estimate of $\mathrm{T^*}$. According to our data, we can determine $\mathrm{T_C}$ $\sim$94 K and $\mathrm{T^*}$ $\sim$170 K, as indicated in Fig. \ref{fig:FigureS1}.

\section{Visible ellipsometry}

\begin{figure}[h!]
\begin{center}
\includegraphics[width=0.7\columnwidth]{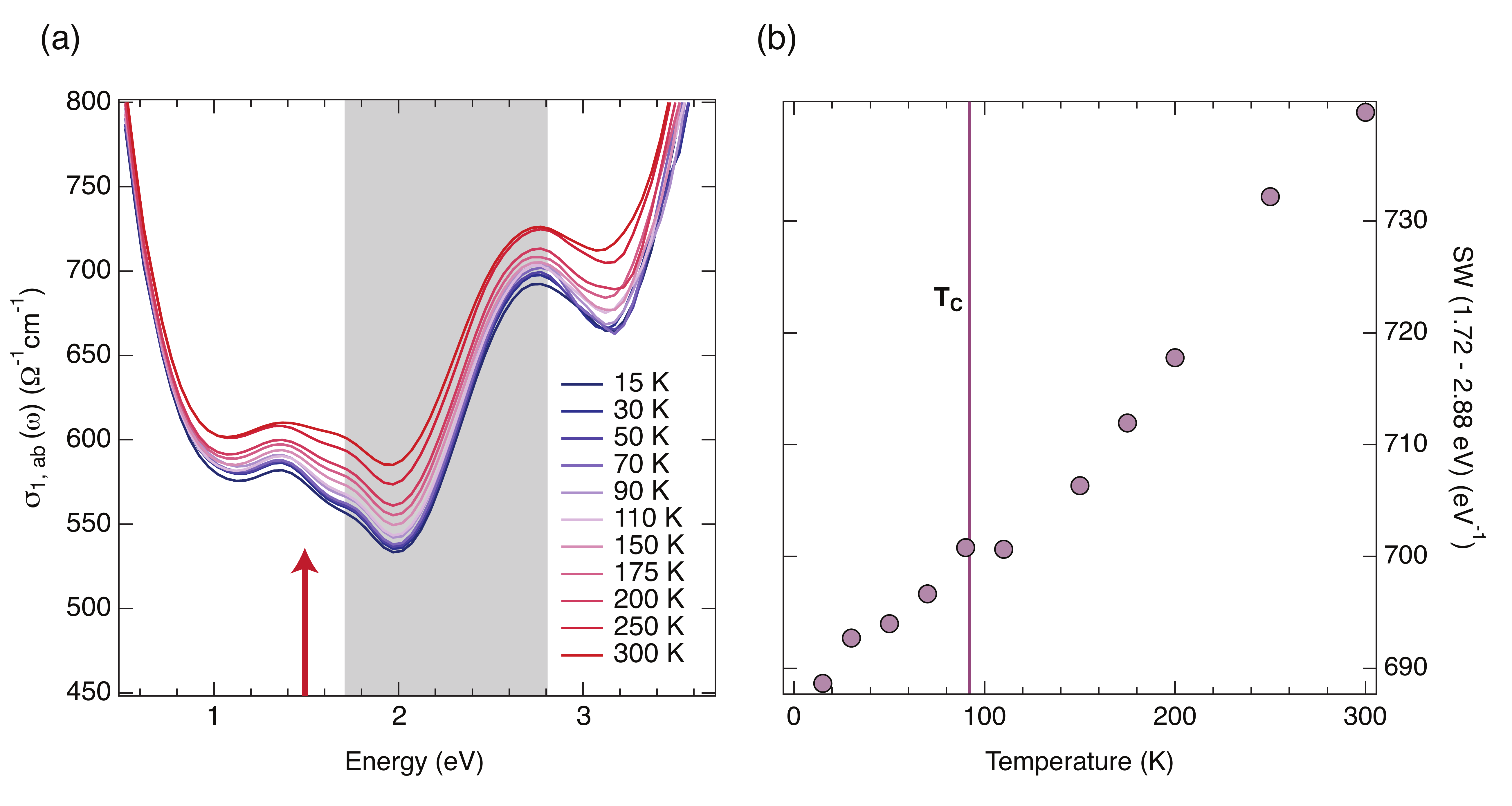}
\caption{(a) Real part of the in-plane optical conductivity at selected temperatures. The red arrow indicates the photon energy of the pump pulse used in the nonequilibrium experiment, while the grey shaded area refers to the spectral range monitored by the broadband probe pulse. (b) Temperature evolution of the spectral weight calculated over the region highlighted by the grey shaded area (1.72 - 2.88 eV). $\mathrm{T_C}$ is indicated on the figure as a violet vertical line.}
\label{fig:FigureS2}
\end{center}
\end{figure}

The in-plane optical conductivity of the sample was also measured by means of broadband spectroscopic ellipsometry from the near-infrared to the ultraviolet. In Fig. \ref{fig:FigureS2}(a), the spectrum of the real part of the optical conductivity $\sigma_{1, ab}(\omega)$ is reported as a function of temperature. The energy of the monochromatic pump (red arrow) and of the broadband probe (grey shaded area), used in the nonequilibrium experiment, are highlighted. The measured response agrees well with previously reported data for the 123 family of cuprates~\cite{ref:cooper, ref:backstrom}. 

The spectral range below 1 eV has been modeled in the past including a Drude-like contribution and a mid-infrared component~\cite{ref:cooper}. The feature at 1.40 eV sharpens when temperature decreases and is closely followed by a weaker satellite at 1.77 eV; it has been ascribed to a charge-transfer excitation in the CuO$_2$ planes and represents a reminiscence of the fundamental absorption gap visible in the same region of the conductivity in the undoped parent compound. Similar features appear in La$\mathrm{_{2-x}}$Sr$\mathrm{_x}$CuO$_4$ and  have been assigned to stripe related bands~\cite{ref:lorenzana}. A second prominent peak can be observed around 2.60 eV, which progressively loses weight with decreasing temperature. Linear Muffin-Tin Orbital calculations assigned this peak to an interband transition into the antibonding Cu(2)-O(2)-O(3) band, with the initial state being found in a manifold of strongly dispersive bands~\cite{ref:kircher}. This interpretation has been recently refined by Dynamical Mean Field Theory calculations, which described it as a charge excitation from the Lower Hubbard Band (LHB) to $\mathrm{E_F}$ in OP YBCO~\cite{ref:fausti}. Figure \ref{fig:FigureS2}(b) shows the partial SW integrated over the region covered by the grey shaded area (1.72 - 2.88 eV) as a function of temperature. We observe that the partial SW drops with decreasing temperature and reveals a kink around $\mathrm{T_C}$. No features are instead seen between $\mathrm{T_C}$ and $\mathrm{T^*}$. The observation of peculiar effects at $\mathrm{T_C}$ taking place in the equilibrium optical spectra at high-energies has been provided by several studies in the past~\cite{ref:backstrom, ref:rubhausen, ref:molegraaf, ref:boris} and interpreted as a fingerprint of Mottness~\cite{ref:stanescu, ref:carbone}.

\begin{figure}[tb]
\begin{center}
\includegraphics[width=0.4\columnwidth]{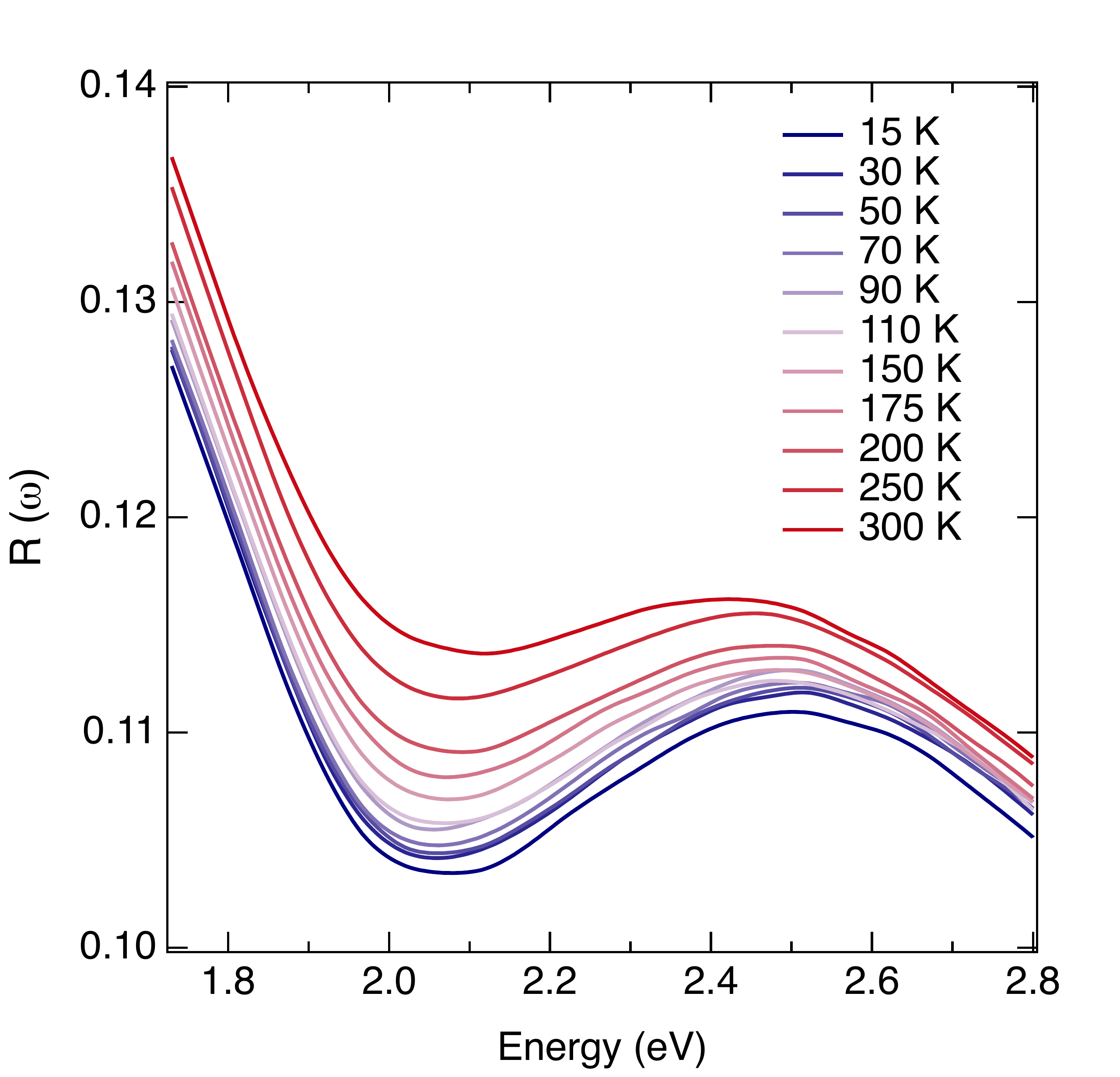}
\caption{Temperature dependence of the in-plane reflectance, R($\omega$), as calculated from the spectroscopic ellipsometry data. The displayed spectral range is limited to the one of the pump-probe experiment.}
\label{fig:FigureS3}
\end{center}
\end{figure}

Finally, we comment on the equilibrium reflectance R($\omega$), that can be calculated from our spectroscopic ellipsometry data. Figure \ref{fig:FigureS3} displays the steady-state reflectance as a function of temperature, limited in the spectral range that is probed by our nonequilibrium experiment (1.73 - 2.80 eV). We observe that, for increasing temperature, R($\omega$) increases its value in the whole spectral range. This confirms that the negative ultrafast response measured by the pump-probe experiment has a nonthermal origin.

\section{Broadband transient reflectivity}

\subsection{Fluence dependence}

\begin{figure}[h!]
\begin{center}
\includegraphics[width=0.5\columnwidth]{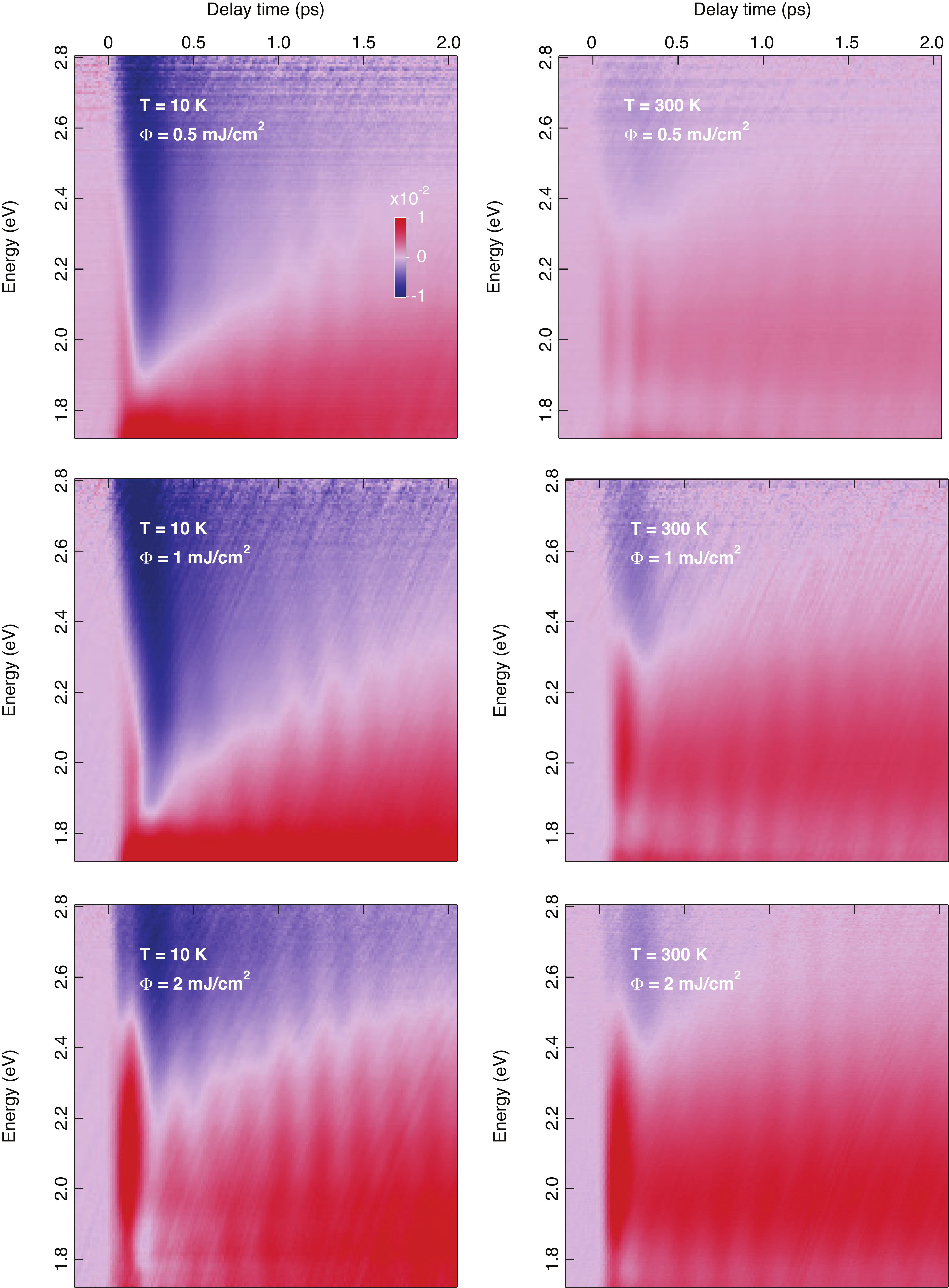}
\caption{Fluence dependence of the transient reflectivity $\Delta$R/R, displayed as a function of the probe photon energy and of the time delay between pump and probe. The temperatures of 10 K and 300 K and the values of the absorbed fluence of 0.5 mJ/cm$^2$, 1 mJ/cm$^2$ and 2 mJ/cm$^2$ are indicated in the labels.}
\label{fig:FigureS4}
\end{center}
\end{figure}

Here, we report the fluence dependence of the in-plane transient reflectivity. Figure \ref{fig:FigureS4} displays $\Delta$R/R($\tau$, E) at 10 K and 300 K for different values of the absorbed fluence (indicated in the labels). At an absorbed fluence of 0.5 mJ/cm$^2$, the low-temperature response resembles the one presented in Fig. \ref{fig:FigureS5}. As a larger number of pump photons are delivered onto the sample, the spectral response of NBCO starts displaying a positive contribution around 2.10 eV and the negative response ascribed to the quasiparticle and the pseudogap contributions vanishes. The rise of this spectroscopic feature represents the signature of a highly thermal nonequilibrium state driven by the pump pulse, which is similar to the signal measured at 300 K.

\subsection{Temperature dependence}

\begin{figure*}[h!]
\begin{center}
\includegraphics[width=0.8\columnwidth]{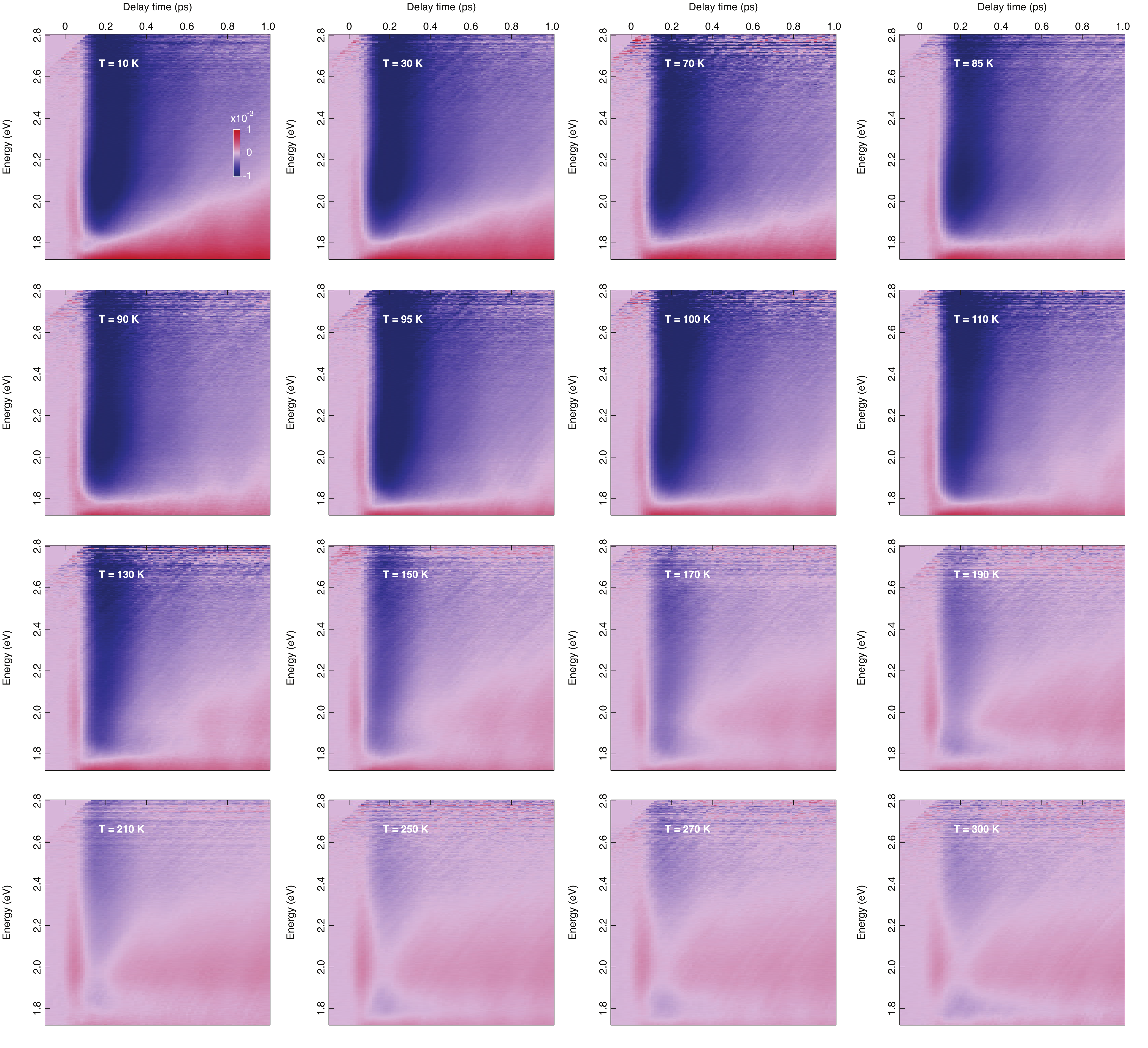}
\caption{Measured temperature dependence of the transient reflectivity $\Delta$R/R($\tau$, E), displayed as a function of the probe photon energy and of the time delay between pump and probe. The absorbed fluence is 0.3 mJ/cm$^2$ and the temperatures are listed as labels in the maps.}
\label{fig:FigureS5}
\end{center}
\end{figure*}

Figure S\ref{fig:FigureS5} reports the complete temperature dependence of the in-plane transient reflectivity $\Delta$R/R as a function of the probe photon energy and the delay time between pump and and probe. The temperatures are reported in the labels.

\subsection{Global fit analysis}

We performed a global fit analysis of $\Delta$R/R as a function of time. Eleven temporal traces have been selected from each map of the temperature dependence and fitted simultaneously by imposing the same time constants. 

\noindent Different models with gradual complexity have been tried to obtain a satisfactory fit of the incoherent response at low temperature. The easiest model which was capable to reproduce the experimental data across the broad spectral range consists of three distinct exponential functions. No converging fit was obtained by using two exponential functions. The function that was used for fitting the data is:

\begin{widetext}
\begin{equation}
f(t) = f_H(t) + f_{PG}(t) + f_{QP}(t) = e^{\frac{-t^2}{\tau_{R_1}}} \ast A_H e^{-\frac{t-t_{D_1}}{\tau_H}} + e^{\frac{-t^2}{\tau_{R_2}}} \ast \Big[
A_{PG} e^{-\frac{t-t_{D_2}}{\tau_{PG}}} + A_{QP} e^{-\frac{t-t_{D_2}}{\tau_{QP}}}\Big]
\end{equation}
\end{widetext}

\noindent where $\mathrm{A_H}$, $\mathrm{A_{PG}}$ and $\mathrm{A_{QP}}$ are the amplitudes of the three exponential functions; $\tau_{R_1}$ and $\tau_{R_2}$ are the rise times of the exponential functions; $\tau_H$, $\tau_{PG}$ and $\tau_{QP}$ are the relaxation constants of the three exponentials; $t_{D_1}$ and $t_{D_2}$ are delay parameters with respect to the zero time.\\

\begin{figure*}[tb]
\begin{center}
\includegraphics[width=\columnwidth]{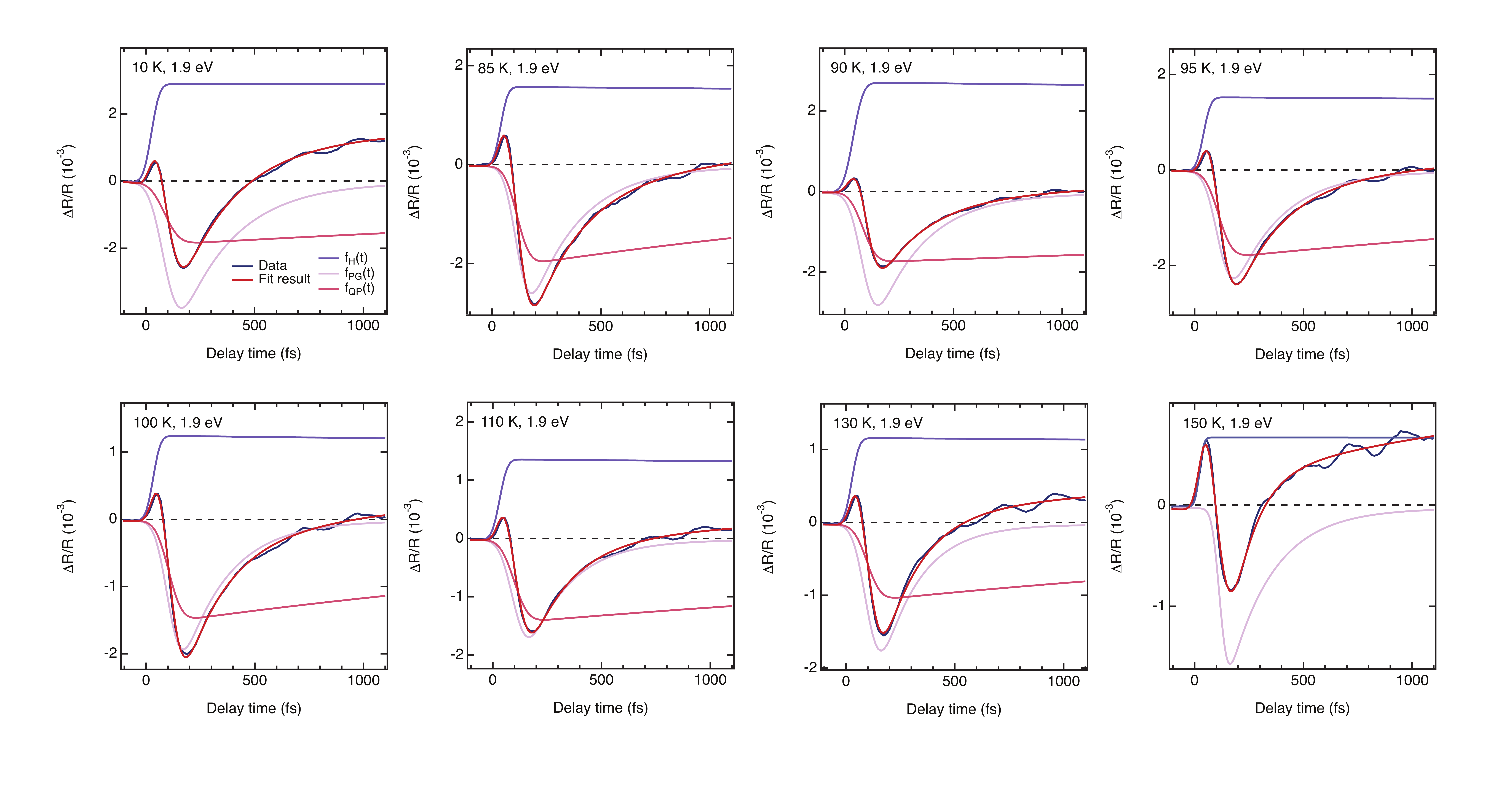}
\caption{Results of the global fit analysis on the temporal trace at 1.9 eV for different temperatures (10, 85, 90, 95, 100, 110, 130, 250 K). The experimental data are reported in blue, the results of the global fit in red. The remaining curves represent the separate contributions to the fitting function.}
\label{fig:FigureS6}
\end{center}
\end{figure*}

\begin{figure*}[tb]
\begin{center}
\includegraphics[width=\columnwidth]{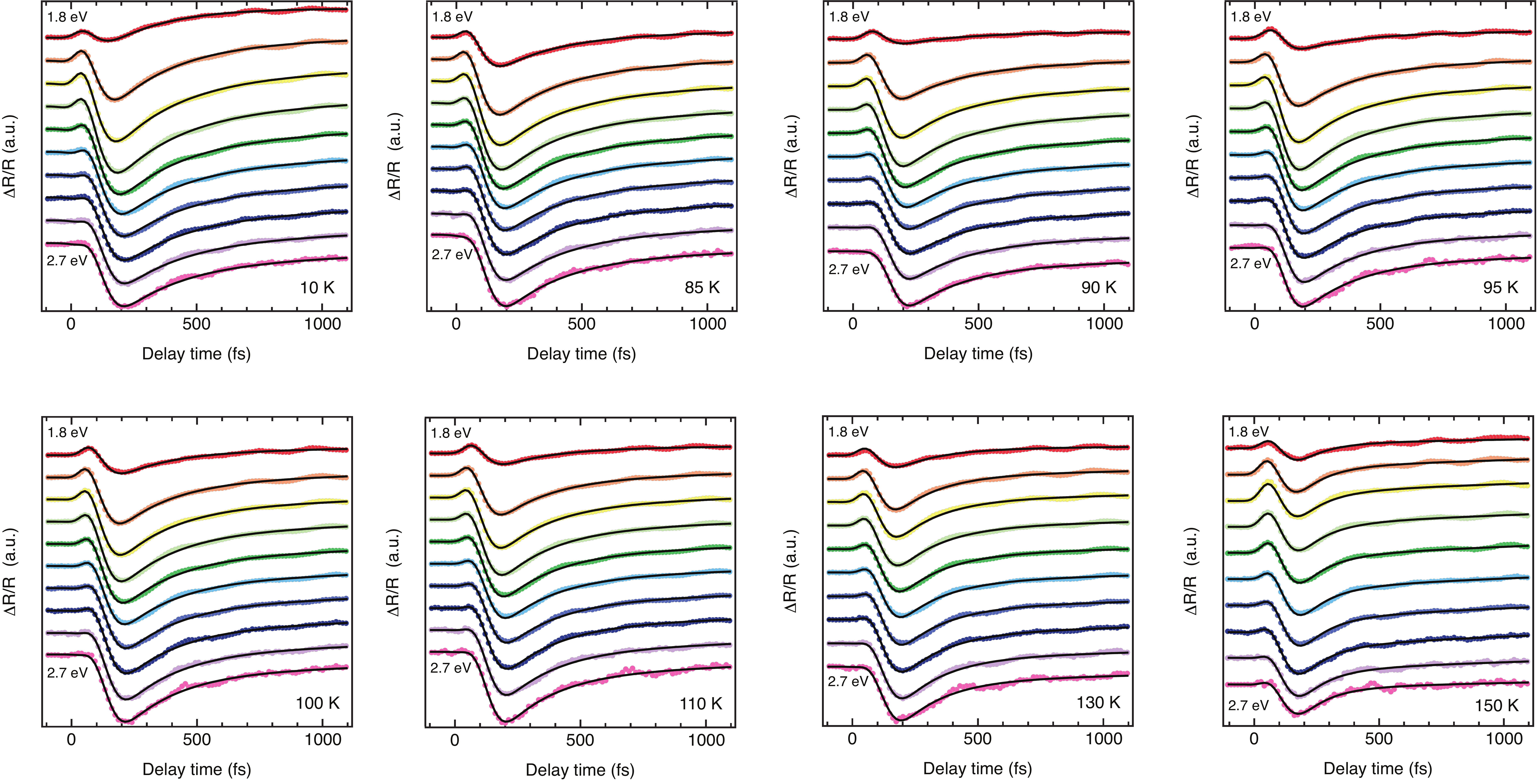}
\caption{Global fit analysis of the temporal traces at 10, 85, 90, 95, 100, 110, 130, 250 K for different probe photon energies (from 1.80 eV at the top of the graphs to 2.70 eV at the bottom, with a constant spacing of 0.10 eV).}
\label{fig:FigureS7}
\end{center}
\end{figure*}

\noindent Depending on the temperature, the extracted values for the time constants varied in the following way:

\noindent $\tau_{PG}$ = 150 $\div$ 250 fs ($\pm$ 4 $\div$ 9 fs), $\tau_{QP}$ = 4 $\div$ 6 ps ($\pm$ 0.65 $\div$ 1.6 ps), $\tau_{H}$ = 34 $\div$ 50 ps ($\pm$ 10 $\div$ 21 ps).

\noindent These three timescales match those reported by all pump-probe experiments on cuprates (see the references cited in the main text). The large uncertainty on the $\tau_{H}$ is due to the limited temporal window probed in our experiment.

\begin{figure}[h!]
\begin{center}
\includegraphics[width=0.6\columnwidth]{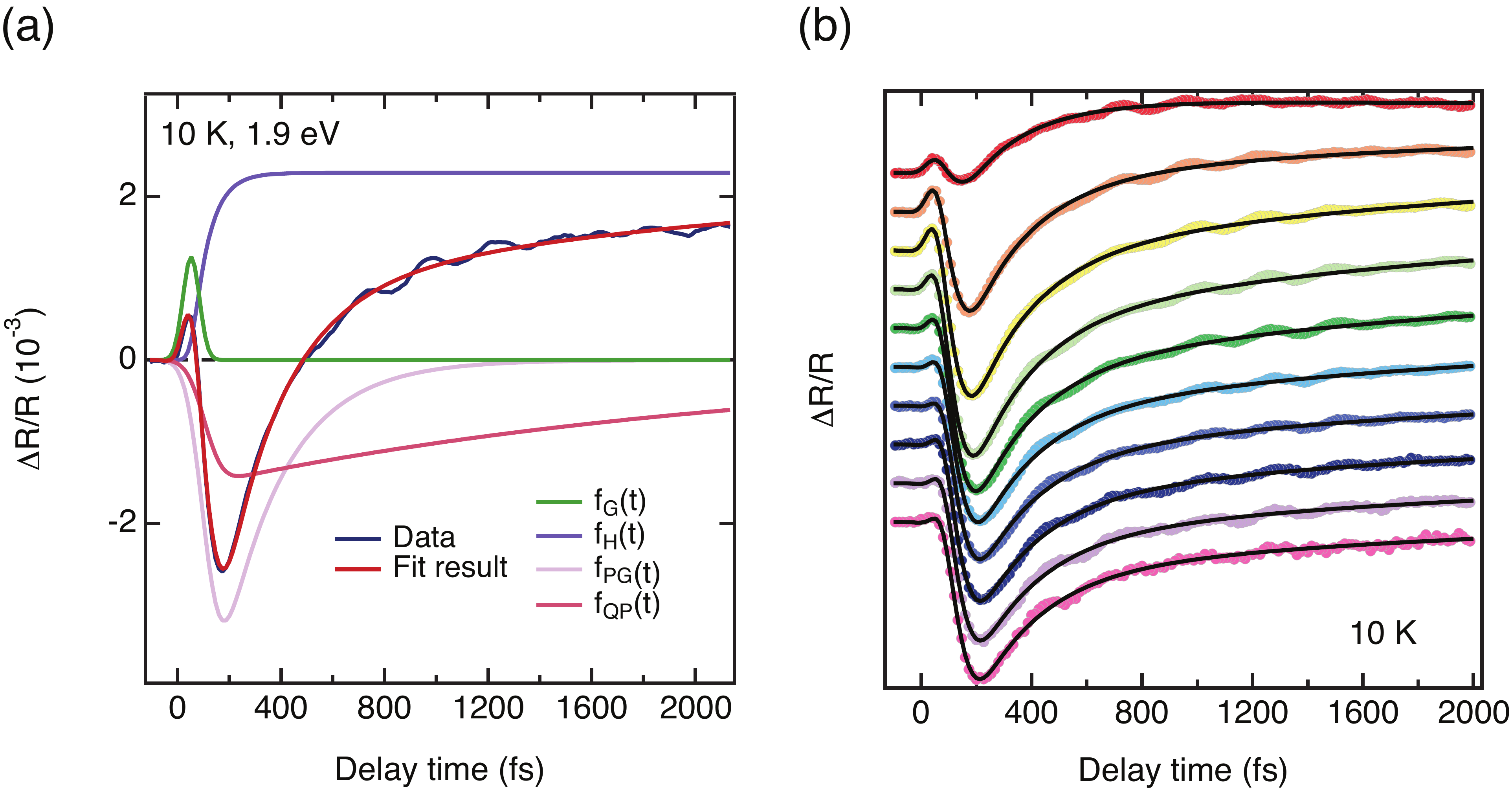}
\caption{(a) Results of the global fit analysis on the temporal trace at 1.90 eV and 10 K. The experimental data are reported in blue, the results of the global fit in red. The remaining curves represent the separate contributions to the fitting function. (b) Global fit analysis of the temporal traces at 10 K for different probe photon energies (from 1.80 eV at the top of the graphs to 2.70 eV at the bottom, with a constant spacing of 0.10 eV).}
\label{fig:FigureS8}
\end{center}
\end{figure}

\begin{figure}[h!]
\begin{center}
\includegraphics[width=0.8\columnwidth]{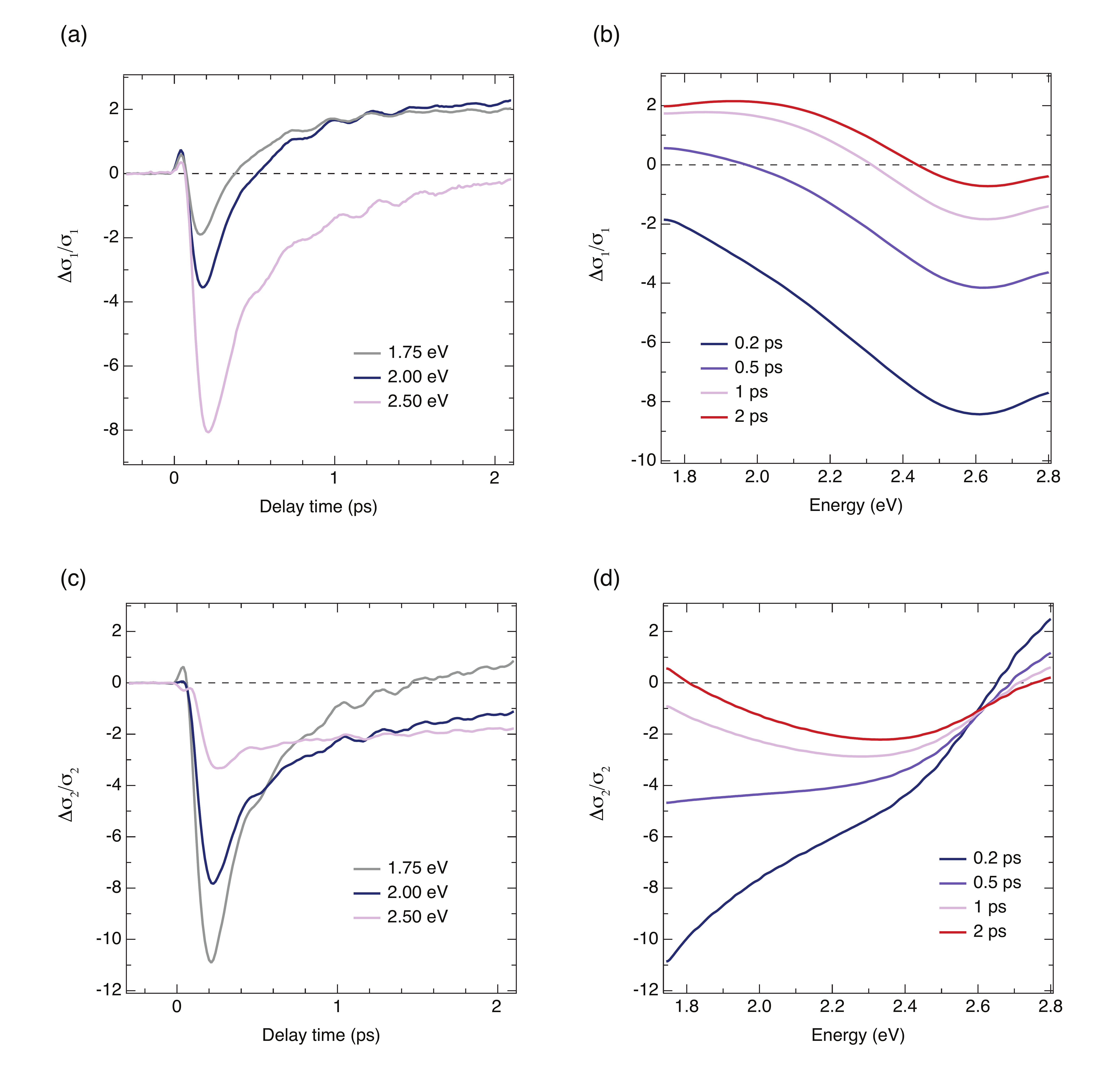}
\caption{(a,c) Temporal traces of $\Delta\sigma_1/\sigma_1$ ($\Delta\sigma_2/\sigma_2$) at 10 K for fixed probe photon energies of 1.75 eV, 2.00 eV and 2.50 eV (b,d) Transient spectrum of $\Delta\sigma_1/\sigma_1$ ($\Delta\sigma_2/\sigma_2$) at 10 K for selected delay times of 0.2, 0.5, 1 and 2 ps.}
\label{fig:FigureS9}
\end{center}
\end{figure}

\noindent The function reported in Eq. (2) allowed to capture both the early temporal dynamics, represented by the resolution-limited positive rise, and the delayed negative contribution whose maximum amplitude occurs around 200 fs. In such a model function, the exponential term with relaxation constant $\tau_H$ embodies the response of localized carriers which remained trapped in a long-lived state~\cite{ref:thomas, ref:cjstevens}; the two exponential terms with delay $t_{D_2}$ represent the conventional PG and QP response reported in all pump-probe spectroscopy experiments. We remark that simpler fit functions (not accounting for the two separate rise times and the two separate delays) could not result in a convergent fit. Hence, although the microscopic explanation for the long-lived component might not be related to an ultrafast localization of charge carriers, we first make use of Eq. (2) for fitting the data in an accurate way, as it represents the simplest fit function with the smallest number of free parameters. Below in this paragraph, we also show that another fit function can be used to successfully reproduce our data at all temperatures, by describing a long-lived bolometric response. At high temperatures, the data were instead fitted using two exponential functions, in agreement with the results obtained in the literature of pump-probe spectroscopy for the analysis of the normal state (which is cited in the main text).

\noindent Figure S6 displays the experimental temporal traces cut around 1.90 eV, together with the results of the global fit analysis. In each panel the result of the fit has been superimposed to the experimental data and the decomposition of the final fit function into its separate contributions is reported as well. The results are shown at different temperatures (10, 85, 90, 95, 100, 110, 130, 250 K). 

\noindent In Fig. S7 we prove the stability of the global fit across the whole probed spectral region. Every panel shows the 11 experimental temporal traces (colored traces) in which the broad spectrum has been cut, together with the results of the global fit (black curves superimposed). The red curve at the top of each graph is the temporal trace at 1.80 eV, while the purple curve at the bottom is the temporal trace at 2.70 eV. We observe that the model used to fit the data is very accurate across the broad spectrum and catches the detailed features characterizing the early dynamics of the transient response.

\begin{figure}[h!]
\begin{center}
\includegraphics[width=0.8\columnwidth]{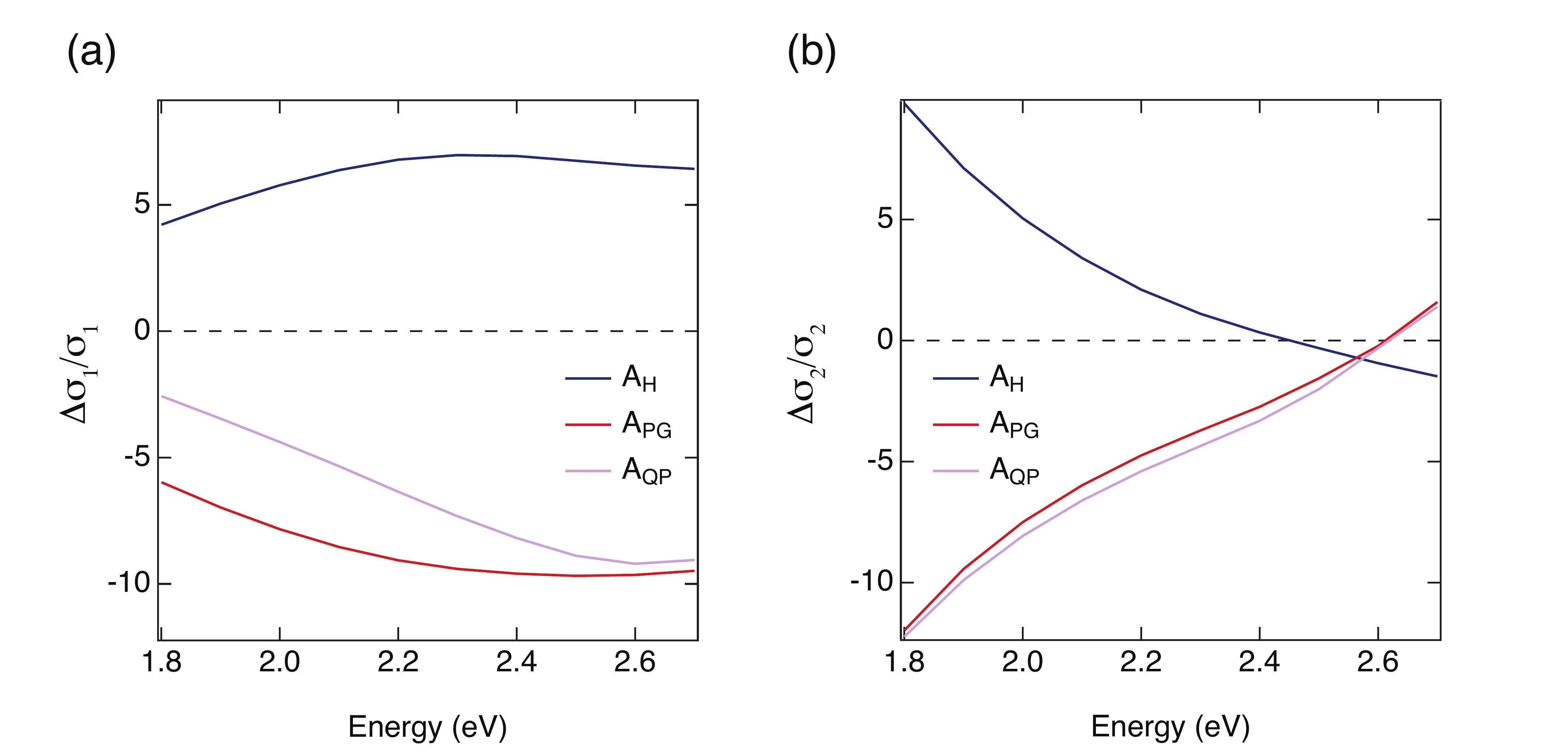}
\caption{Transient optical conductivity spectra of the three components obtained from the global fit analysis. (a) $\Delta\sigma_1/\sigma_1$; (b) $\Delta\sigma_2/\sigma_2$.}
\label{fig:FigureS9}
\end{center}
\end{figure}

\noindent As anticipated above, we also used a second approach for fitting the experimental data, which involves a Gaussian pulse-like contribution for capturing the early dynamics, a slowly rising long-lived component and the two delayed exponential functions of the previous model. The separate contributions are depicted in Fig. S8(a) for the temporal trace at 1.90 eV at 10 K. In this case, the Gaussian pulse is used to model the initial metallic contribution setting after the interaction of the pump with the sample, following the model of Okamoto \textit{et al.}~\cite{ref:okamoto}. The delayed negative exponential functions still embody the PG and QP response, decaying with time constants $\tau_{PG}$ and $\tau_{QP}$. The long-lived component instead corresponds to the bolometric (heating) response of the sample, which sets after the thermalization of the excited carriers has occurred. Although the use of this fit function led to a satisfactory description of the experimental data, the number of free parameters was larger than the function proposed in the previous model. In any case, the use of this second fitting approach did not change the outcome of our analysis in a substantial way (Fig. S8(b)).

\subsection{Transient optical conductivity}

We calculated the transient optical conductivity $\Delta\sigma_1/\sigma_1$ from the transient reflectivity $\Delta$R/R following the procedure reported in the main text. Figures S9(a)-(b) display the temporal traces and the transient spectrum of $\Delta\sigma_1/\sigma_1$ at 10 K, which have been cut from the map displayed in Fig. 5 of the main text. Figures S9(c)-(d) shoe the corresponding temporal traces and the transient spectrum of $\Delta\sigma_2/\sigma_2$.

\begin{figure}[h!]
\begin{center}
\includegraphics[width=0.8\columnwidth]{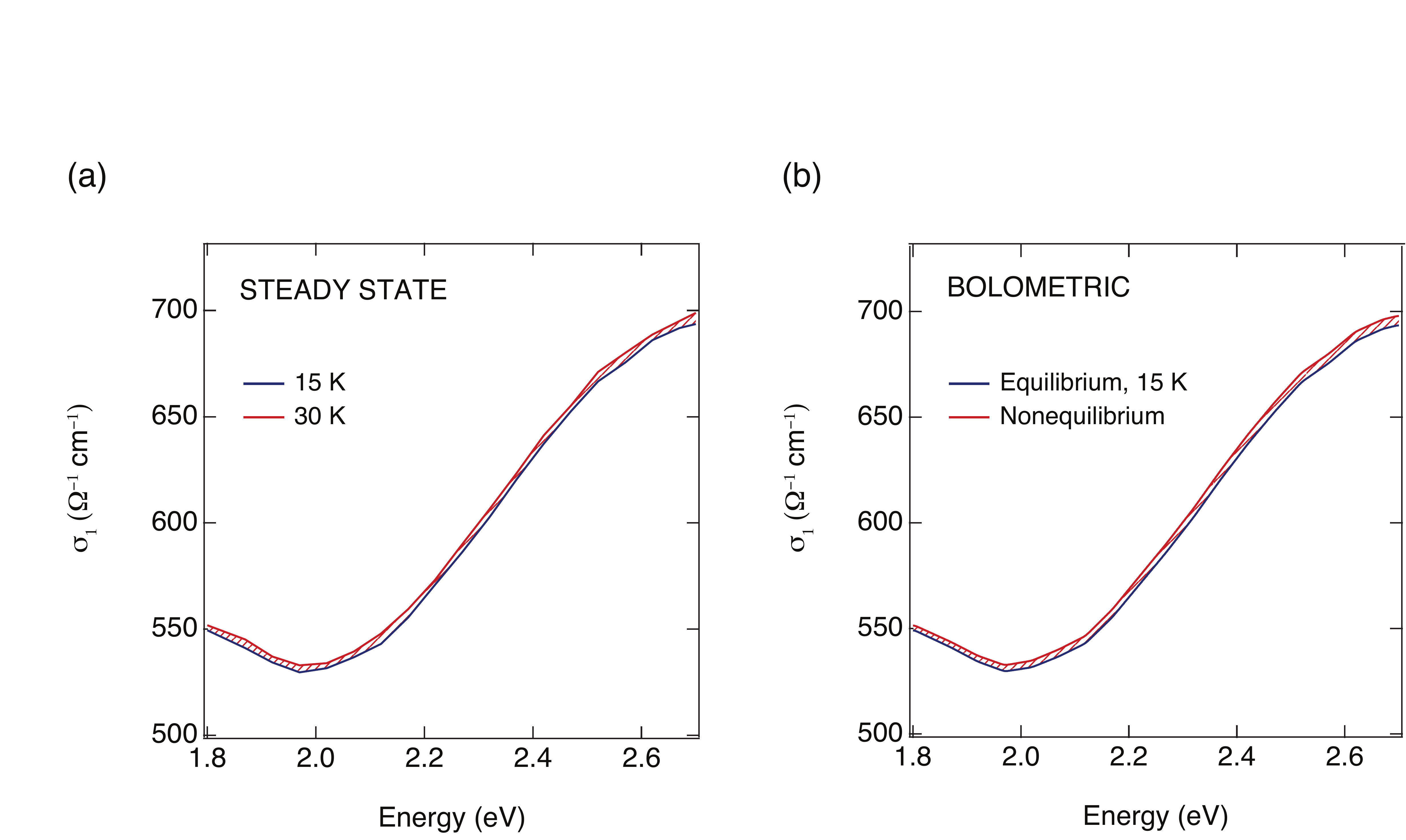}
\caption{(a) Steady-state optical conductivity measured by spectroscopic ellipsometry in the visible range at 15 K (blue curve) and 30 K (red curve). (b) Effect produced on the equilibrium $\sigma_1$ (represented in blue) by the long-lived component in the transient signal. This can be identified as the bolometric response of the crystal after photoexcitation.}
\label{fig:FigureS11}
\end{center}
\end{figure}

\begin{figure}[h!]
\begin{center}
\includegraphics[width=\columnwidth]{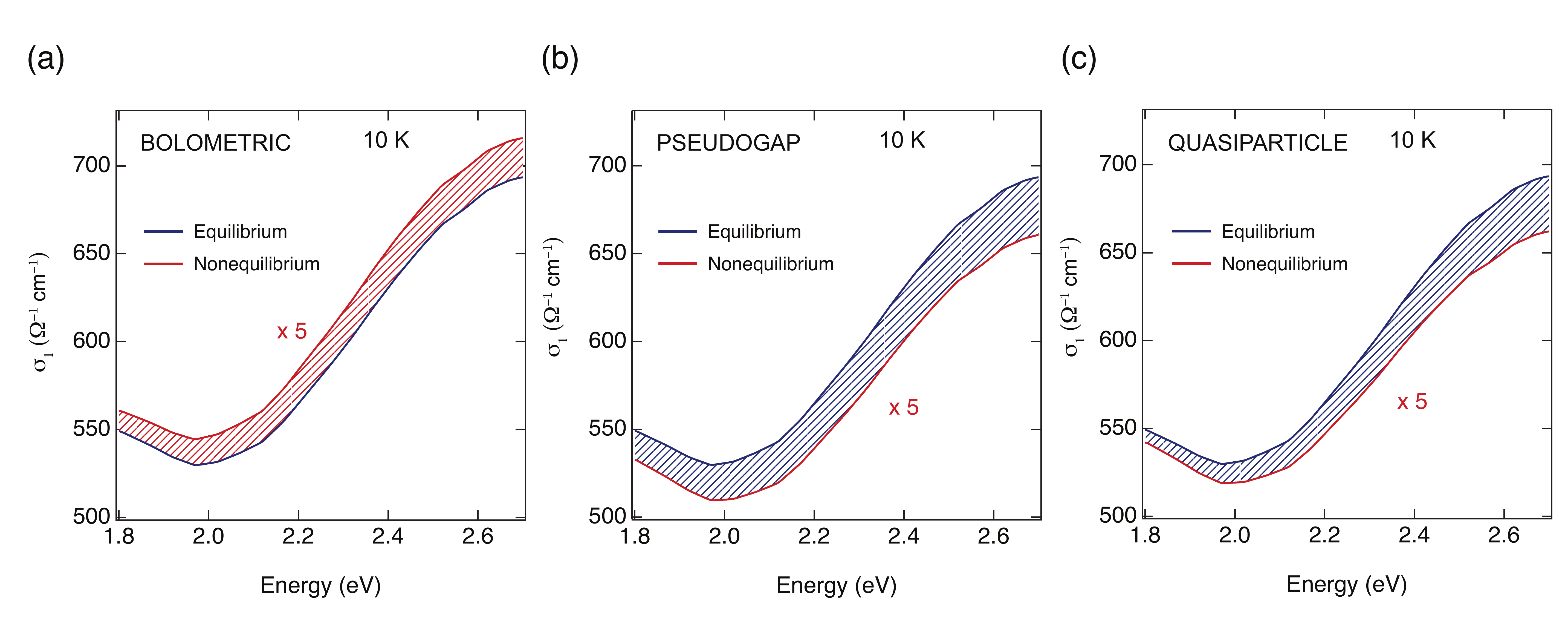}
\caption{Effects produced on the equilibrium $\sigma_1$ (represented as blue lines) by (a) the bolometric, (b) the PG and (c) the QP responses. The nonequilibrium $\sigma_1$ is drawn as a red curve and the added (removed) SW is depicted by the red (blue) filled area.}
\label{fig:FigureS12}
\end{center}
\end{figure}

\noindent We selected ten temporal traces across the broadband spectral range and we performed a global fit analysis, by imposing the same model used for the global fit of $\Delta$R/R. This global fit allowed us to isolate the spectral fingerprints of the three separate components emerging in our nonequilibrium experiment. These spectral components are shown in Fig. S10(a)-(b), in which we used the same nomenclature reported in Eq. (2). We observe that $\mathrm{A_H}$ provides a positive contribution to $\Delta\sigma_1/\sigma_1$, while $\mathrm{A_{PG}}$ and $\mathrm{A_{QP}}$ are at the origin of the decreased $\Delta\sigma_1/\sigma_1$ which is dominating in the traces and spectra of Fig. S9(a).

\noindent Here, we can contribute to the discussion concerning the long-live component, which has been previously assigned either to a pump-induced heating effect or as the signature of an excited-state absorption from localized (polaronic) carriers~\cite{ref:thomas, ref:cjstevens}. Indeed, in Fig. S11 we compare the effect produced on $\sigma_1$ by this long-lived component in the nonequilibrium experiment and the variation in $\sigma_1$ provided by a static increase of the crystal temperature (from 10 to 30 K). We observe an excellent matching between the two trends, which suggests that the long-lived component corresponds to the bolometric response of our sample. Notice that this assignment implies that the correct model to fit the data is the one described in Fig. S8(a), as the rise of the bolometric response is strictly correlated to the thermalization of the hot charge carriers after the initial excitation.

\noindent Finally, in Fig. S12, we report the effects produced by each separate component on the equilibrium $\sigma_1$. In each panel, the variation occurring in $\sigma_1$ has been multiplied by a factor 5 to make the change distinguishable. As explained in the main text, both the PG and QP responses remove SW in the visible region, redistributing it most likely to lower energy.

\subsection{Coherent phonons}

We also tracked the amplitude of the Ba and Cu modes as a function of the probe photon energy, by perforing a Fourier transform analysis of the residuals from the global fit. Thanks to the broadband probe, this methodology allows to map the Raman matrix element for each Raman-active mode which is coherently excited in the pump-probe experiment \cite{ref:fausti, ref:mann}. In Fig. S13 we report the separate Raman matrix elements for the Ba (blue curve) and Cu (red curve) modes at 10 K. 

\begin{figure}[h!]
\begin{center}
\includegraphics[width=0.35\columnwidth]{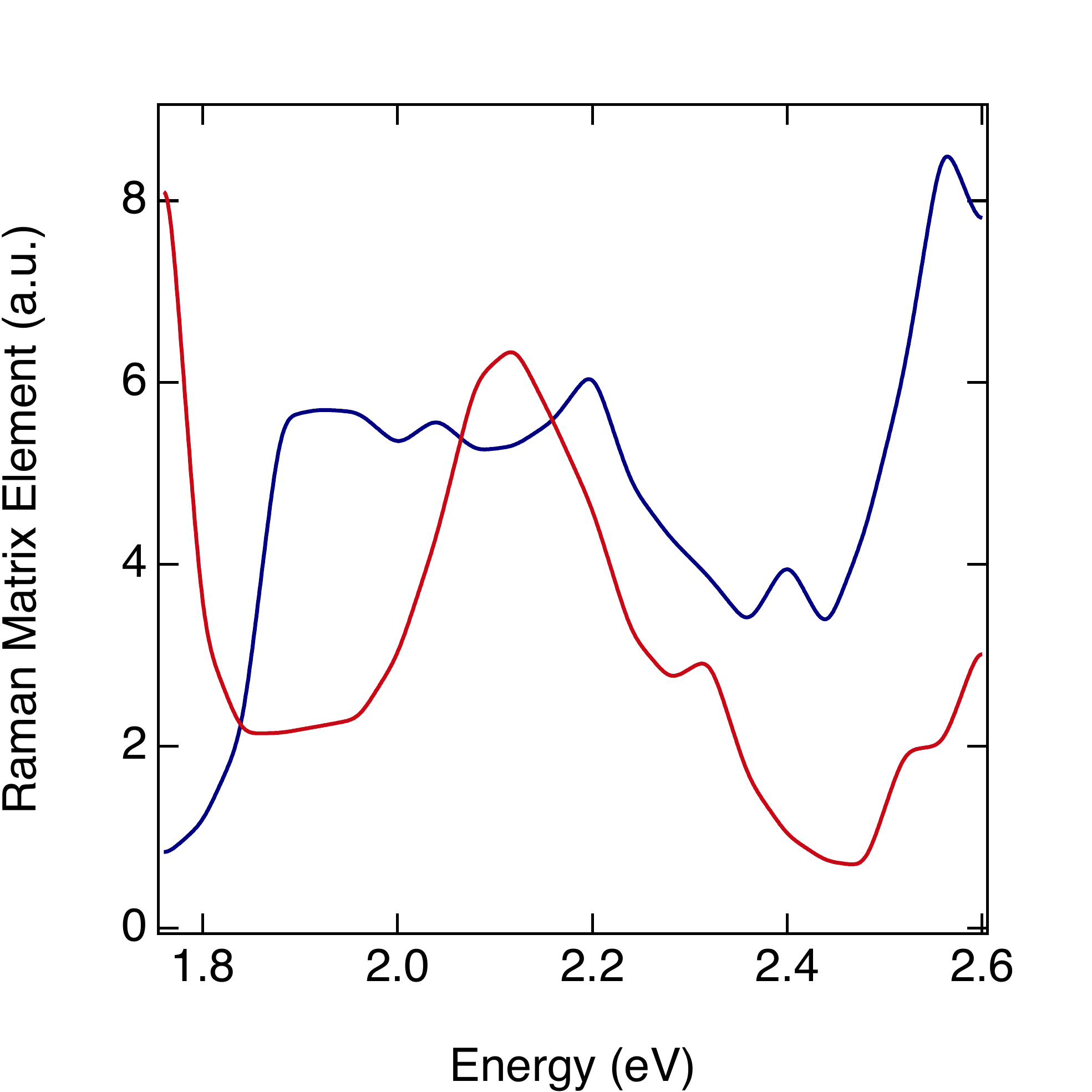}
\caption{Raman matrix elements of the Ba (blue curve) and Cu (red curve) modes at 10 K, as determined from the Fourier transform analysis as a function of the probe photon energy.}
\label{fig:FigureS12}
\end{center}
\end{figure}

\noindent We observe that at 2.00 eV the Ba mode intensity is larger than the one of Cu mode by a factor 1.7. This agrees with the results reported by single-wavelength pump-probe spectroscopy at 2.00 eV, in which the low-temperature signal was found to be dominated by the contribution of the Ba mode~\cite{ref:albrecht}. The general trend described by our Raman matrix elements also agrees with early spontaneous Raman scattering experiments performed at varying laser photon energies~\cite{ref:friedl}. Finally, it is worth to underline the resonant behaviour displayed by the Cu mode for probe photon energies below 1.80 eV, \textit{i.e.} outside our probed spectral range. This finds a possible explanation in the proximity of the spectroscopic feature at 1.40 - 1.77 eV (Fig. S2(a)), which has been associated with the reminescence of the charge-transfer excitation in the CuO$_2$ planes. We remark that the shape of the determined Raman matrix elements are very similar to the ones reported in Ref.~\cite{ref:fausti} for OP YBCO.
\newpage

\bibliography{Papers}
\end{document}